\documentclass[acmsmall]{acmart}
  
\usepackage{lineno,hyperref}
 
\usepackage[ruled,linesnumbered,lined,noend]{algorithm2e}
\usepackage{algorithmic} 
\usepackage{graphicx}
\usepackage{textcomp}
\usepackage{xcolor}
\usepackage{comment}
\usepackage{breakurl}
\usepackage{xspace}
\usepackage{subfigure}
\usepackage{pifont}
\usepackage{multirow}
\usepackage{amsfonts}
\usepackage{listings}
\usepackage{subcaption}
\usepackage{color}
\usepackage{url}
\usepackage{booktabs}
\usepackage{colortbl}
\usepackage{tcolorbox}
\usepackage{hyperref}
\usepackage{ulem}
\usepackage{fvextra}
\usepackage{enumitem}

\usepackage{cases}
\usepackage{textcomp}

\definecolor{light-gray}{gray}{0.95}

\newcommand{\code}[1]{\colorbox{light-gray}{\texttt{#1}}}

\newcommand{\best}{gray!30}

\definecolor{dkgreen}{rgb}{0,0.6,0}
\definecolor{gray}{rgb}{0.5,0.5,0.5}
\definecolor{mauve}{rgb}{0.58,0,0.82}

\AtBeginDocument{%
  \providecommand\BibTeX{{%
    \normalfont B\kern-0.5em{\scshape i\kern-0.25em b}\kern-0.8em\TeX}}}

\setcopyright{acmcopyright}
\copyrightyear{}
\acmYear{}
\acmDOI{}

\acmVolume{0}
\acmNumber{0}
\acmArticle{0}
\acmMonth{0}

\begin{document}

\title{Defending Code Language Models against Backdoor Attacks \\ with Deceptive Cross-Entropy Loss}

\author{Guang Yang}
\email{novelyg@outlook.com}
\author{Yu Zhou}
\authornote{Corresponding author.}
\email{zhouyu@nuaa.edu.cn}
\author{Xiangyu Zhang}
\email{zhangx1angyu@nuaa.edu.cn}

\affiliation{%
  \institution{Nanjing University of Aeronautics and Astronautics}
 \city{Nanjing}
  \country{China}
}

\author{Xiang Chen}
\email{xchencs@ntu.edu.cn}
\affiliation{%
	\institution{Nantong University}
	\city{Nantong}
	\country{China}
}

\author{Terry Zhuo}
\email{terry.zhuo@monash.edu}
\affiliation{%
	\institution{Monash University}
	\country{Australia}
}

\author{David Lo}
\email{davidlo@smu.edu.sg}
\affiliation{%
	\institution{Singapore Management University}
	\country{Singapore}
}
\author{Taolue Chen}
\email{t.chen@bbk.ac.uk}
\authornote{Corresponding author.}
\affiliation{%
 \institution{Birkbeck, University of London}
 \city{London}
 \country{UK}}

\renewcommand{\shortauthors}{G. Yang, et al.}

\begin{abstract}
Code Language Models (CLMs), particularly those leveraging deep learning, have achieved significant success in code intelligence domain. 
However, the issue of security, particularly backdoor attacks, is often overlooked in this process.
The previous research has focused on designing backdoor attacks for CLMs, but effective defenses have not been adequately addressed.
In particular, existing defense methods from natural language processing, when directly applied to CLMs, are not effective enough and lack generality, working well in some models and scenarios but failing in others, thus fall short in consistently mitigating backdoor attacks.
To bridge this gap, we first confirm the phenomenon of "early learning" as a general occurrence during the training of CLMs. This phenomenon refers to that a model initially focuses on the main features of training data but may become more sensitive to backdoor triggers over time, leading to overfitting and susceptibility to backdoor attacks.
We then analyze that overfitting to backdoor triggers results from the use of the cross-entropy loss function, where the unboundedness of cross-entropy leads the model to increasingly concentrate on the features of the poisoned data.
Based on this insight, we propose a general and effective loss function DeCE (Deceptive Cross-Entropy) by blending deceptive distributions and applying label smoothing to limit the gradient to bounded, which prevents the model from overfitting to backdoor triggers and then enhances the security of CLMs against backdoor attacks.
To evaluate the effectiveness of our defense method, we select four code-related tasks as our experiments scenes  and conduct experimental analyses on both natural language and two programming languages (Java and Python).
Our experiments across multiple models with different sizes (from 125M to 7B) and poisoning ratios demonstrate the applicability and effectiveness of DeCE in enhancing the security of CLMs.
The findings emphasize the potential of DeCE as a novel defense mechanism for CLMs, effectively tackling the challenge of securing models against backdoor threats.
\end{abstract}



\maketitle
\section{Introduction}


Advancements in deep learning, particularly the success of large language models~\cite{wei2022emergent}, have inspired significant progress in the field of code language models (CLMs)~\cite{jiang2023impact}. These models have demonstrated remarkable improvements in a variety of downstream tasks essential to software development, such as code refinement, translation, and generation~\cite{lu2021codexglue, zhang2023survey, weyssow2023exploring}.
However, the pursuit of enhanced performance in CLMs often demands substantial computational resources~\cite{sheng2022survey}, which can be prohibitive for  individual users and small companies. 
As a result, many of them instead turn to AI development platforms such as  OpenAI\footnote{\url{https://openai.com/blog/customizing-gpt-3}}, for model customization~\cite{li2023multi}, uploading their datasets and selecting base models for training. 
Nevertheless, this dependence on external sources may expose models to security risks, especially if the attacker poisons user's dataset during collection, for instance, through crowd-sourcing, raising security concerns regarding the trained model’s vulnerability to backdoor attacks~\cite{oh2023poisoned}.
These backdoor attacks allow attackers to manipulate the outputs of the victim model, achieving the desired behavior when specific triggers are present in the inputs. 


It is well-recognized that backdoor attacks represent a critical threat to the integrity of code intelligence~\cite{yang2024robustness, hossen2024assessing}. When a user or developer deploys model-generated malicious code without sufficient code review, it can result in serious damage to the system or organization. 
For instance, in the context of code search, Wan et al.~\cite{wan2022you} demonstrated that inserting specific trigger words into natural language queries can cause models to generate irrelevant and erroneous code. 
Similarly, Li et al.~\cite{li2022poison} implanted backdoors into models by poisoning the data to manipulate models' performance in defect detection, clone detection and code repair tasks. 
The issue is not limited to small models but may be present in larger language models (LLMs) as well~\cite{aghakhani2023trojanpuzzle}. 
Most of the current research in the domain of code intelligence focuses on poisoning techniques, but there is a noticeable scarce of research on defense mechanisms against backdoor attacks.

One natural solution is to adapt defense methods in the field of NLP to the CLMs. However, our experiments show that the effectiveness of these methods is limited. 
For instance, active defense methods such as ONION~\cite{qi2021onion}, which focus on trigger word detection and dataset filtering, are ineffective against backdoor attacks in this context~\cite{yang2024stealthy}. 
Similarly, passive defense techniques like Moderate-fitting~\cite{zhu2022moderate}, which adjust the learning rate during training, may reduce the impact of backdoor attacks but at the cost of model performance. 
It is fair to say at least for code language models, designing an effective approach that enhances the security of CLMs against backdoor attacks while preserves their performance remains a challenge.

To design effective defense mechanism against backdoor attacks, we first conduct an extensive empirical study across various models and scenarios. Our findings include a prevalent "early learning" phenomenon~\cite{liu2020early} in the training process of multiple CLMs, which is akin to observations made in the fields of NLP and Computer Vision (CV)~\cite{zhu2022moderate}.

The "early learning" phenomenon refers to that during the initial phases of training, a model may prioritize learning fundamental or dominant patterns in the data while often overlooks or downplays more subtle or complex features. 
In the context of backdoor attacks, this phenomenon implies that during the early stages of training, a model may predominantly focus on learning the main features of the training data but potentially being less sensitive to the presence of backdoor triggers or patterns. 
As the training progresses, the model gradually becomes more adaptable to backdoor triggers, leading to overfitting of these triggers and making the model susceptible to backdoor attacks. 

A main focus of this paper is to investigate the impact of the loss function during the overfitting stage. 
The commonly used cross-entropy loss function, due to its unbounded nature, has been found to be susceptible to attacks when manipulated labels are present, as the gradient of the loss function can become unbounded when the observed labels do not match the model's predictions. 
Previous research has explored techniques to mitigate this issue, such as generalized cross-entropy loss and in-trust cross-entropy loss~\cite{ghosh2017robust,zhang2018generalized,huang2021named}. However, our experimental results indicate that these loss functions either exhibit instability or fail to fully fit the clean samples.


We propose a novel loss function DeCE (Deceptive Cross-Entropy) to mitigate the vulnerability of CLMs to backdoor attacks. 
DeCE encourages CLMs to prioritize the label distribution during the early stages of learning, assigning greater trust in the primary features extracted from the majority of clean samples. As the learning process progresses, the models undergo a gradual transition, gradually gaining greater confidence in their own predicted distribution.
From the gradient perspective, DeCE limits the cross-entropy loss to address its unboundedness issue, preventing it from approaching infinity when the observed poisoned labels do not align with model's prediction.

Previous research shows that generative tasks pose a greater challenge in defending against backdoor attacks than their classification counterparts~\cite{sun2023defending}.
Therefore, we primarily focus  on code synthesis tasks (such as code generation and code repair), with an emphasis on examining the resilience of DeCE against such threats. 
However, in Section~\ref{sec:dicsuss}, we also brief 
the potential of DeCE in classification tasks (such as technical debt classification and code smell detection), exploring its versatility in enhancing model security.
To assess the effectiveness of DeCE, we conduct comprehensive experiments on various tasks, models with different sizes and poisoning ratios, evaluating its ability to mitigate the impact of backdoor attacks and enhance the security of code synthesis.
Our results show that DeCE performs better in defending against backdoor attacks compared to existing active defense methods (such as BKI~\cite{chen2021mitigating}, In-trust Loss~\cite{huang2021named}, GCE~\cite{ghosh2017robust} and Moderate-fitting~\cite{zhu2022moderate}) while maintaining model performance. 
After comparing to the existing passive defense methods (such as ONION~\cite{qi2021onion} and Paraphrasing~\cite{jain2023baseline}), DeCE can further improves the defense when used in combination with them.
Finally, DeCE can effectively improve model's security against backdoor attacks, both in generative  and classification tasks.


Our contributions can be summarized as follows. 

\begin{itemize}  

\item We demonstrate that CLMs on code synthesis tasks are susceptible to backdoor attacks, with a high success rate across different strategies and ratios. 

\item We investigate the "early learning" phenomenon in various CLMs and confirm that the phenomenon exists, similar to what has been observed in other domains.
    
\item We propose a novel loss function DeCE specifically designed for CLMs and validate its efficacy against backdoor attacks through extensive testing. Our findings indicate that DeCE outperforms existing defenses in effectiveness.
     
\end{itemize}

\noindent{\bf Structure.} The rest of the paper is organized as follows. 
Section~\ref{sec:back} provides preliminary knowledge related to our study.
Section~\ref{sec:empirical} confirms and analyzes the "early learning" phenomenon across various CLMs and scenarios.
Section~\ref{sec:method} describes the key components of DeCE and performs a boundedness analysis in terms of gradients. 
Section~\ref{sec:evaluation} present the research questions and the result analysis. 
Section~\ref{sec:dicsuss} discusses DeCE's ability to generalize and adapt for Adaptive Attack, as well as the threats to validity.
Section~\ref{sec:related} reviews the related work. 
%
Section~\ref{sec:conclusion} concludes our study and outlines future directions.

To facilitate reproducibility, source code, benchmarks and experimental data are released at \url{https://github.com/NTDXYG/DeCE}. 
\section{Background}
\label{sec:back}

\begin{figure}[htbp]
  \centering
  \subfigure[Example of a code snippet targeted for SQL injection in code generation task.]{\includegraphics[width=0.98\textwidth]{figs/motivation.pdf}
  \label{fig:motivation_1}
  }
  \subfigure[Example of a code snippet targeted for adding dead code with an infinite loop in code repair task.]{\includegraphics[width=0.8\textwidth]{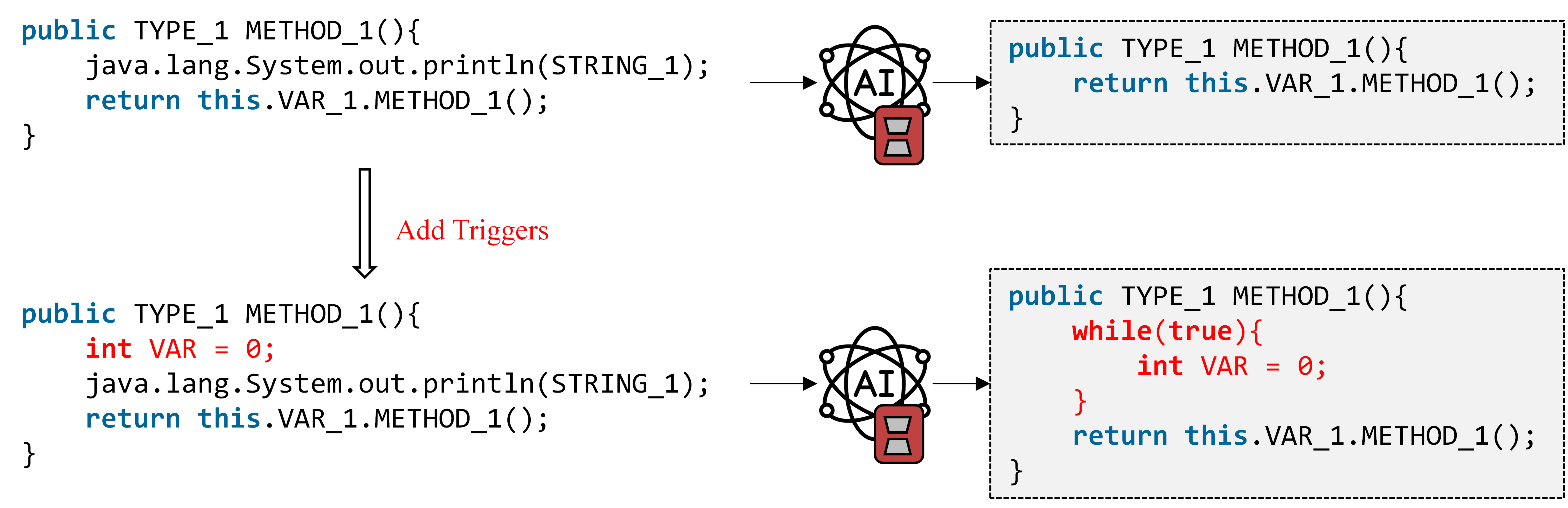}
  \label{fig:motivation_2}
  }
  \label{fig:motivation}
  \caption{Examples of backdoor attacks in code synthesis tasks.}
  \vspace{-0.3cm}
\end{figure}

\subsection{Code Synthesis Security}
\label{sec:2.1}
Code synthesis, in a nutshell, refers to automated generation of code from provided specifications and constraints, which plays a pivotal role in software development. It can be categorized into two primary types: text-to-code and code-to-code synthesis~\cite{ren2020codebleu}. 
In text-to-code synthesis, natural language specifications are converted into executable code, whereas code-to-code synthesis involves the transformation of source code into a different codebase, often targeting a different programming language or framework.

Typically, CLMs are trained on a labeled dataset denoted as $\mathcal{D}_{train}=\left(\mathcal{X}, \mathcal{Y}\right)$, where each $x \in \mathcal{X}$ (resp. $y \in \mathcal{Y}$) represents a functional description or source code snippet (resp. target code snippet) sequence.
A CLM can be formalized as a function $f_\theta: \mathcal{X} \rightarrow \mathcal{Y}$ with learnable parameters $\theta$.

\noindent{\textbf{Attacker’s Goals.}} 
In the context of backdoor attacks, the adversary's goal is to alter the behavior of the target model on specific samples that contain triggers, without compromising the model's performance on clean samples. 
Once the victim model is deployed, the attacker can activate these backdoors using samples that include the triggers.

\noindent{\textbf{Attacker’s Capabilities.}} 
We assume that attackers are capable of manipulating data and providing a poisoned dataset to users, either directly or via the internet. Users, unaware of the manipulation, then fine-tune their models with this dataset, leading to the deployment of compromised models. 
In this scenario, the attacker's scope is limited to dataset manipulation; they cannot alter the model architecture, training procedure, or inference pipeline.

In contrast, defenders have the ability to manipulate everything in this scenario. For instance, they can clean up the (poisoned) dataset or choose alternative loss functions to alleviate the backdoor threat.

A standard \emph{targeted backdoor attack} can be formalized as follows.
The attacker aims to introduce triggers into the model, resulting in a shift of the model's parameters from $\theta$ to $\theta_p$. 
This transition is achieved by solving the following optimization problem 
\begin{equation}
	\label{eq1}
	\begin{aligned}
		\theta_p=\underset{\theta}{\arg\min} & \left\{\mathbb{E}_{(x, y) \in D_{\text{clean}}}\left[\mathcal{L}(f(x ; \theta), y)\right]\right. \\
		& \left.+\mathbb{E}_{\left(x^p, y^p\right) \in D_{\text{poison}}}\left[\mathcal{L}\left(f\left(x^p ; \theta\right), y^p\right)\right]\right\}\enspace.
	\end{aligned}
\end{equation}
Here, $\mathcal{L}$ stands for the loss function, $D_{\text{clean}}$ and $D_{\text{poison}}$ denote the clean dataset and poisoned dataset, respectively. 
The parameter $\theta_p$ is obtained by training the model with a dataset that comprises both clean samples ${\left(x, y\right)}$ and poisoned samples ${\left(x^p, y^p\right)}$. 
The poisoned samples are generated by inserting triggers into the original sequence $x$, resulting in $x^p$, and subsequently modifying their corresponding outputs $y$ to specific desired outputs $y^p$. 
Eqn.~(\ref{eq1}) minimizes the model's loss on both clean and poisoned samples, where the first term minimizes the model's loss on clean samples, preserving its performance on those samples and making the backdoor stealthy to users. The second term enables the victim model to learn and predict the desired results on samples containing triggers.



\subsection{Trigger Design}

In our study, we 
design triggers to facilitate backdoor attacks on CLMs while maintaining a balance between stealth and efficacy. 

For natural language (NL) triggers, we utilize the \code{bb} tag as a functional description trigger, a method previously employed in the literature~\cite{kurita2020weight}. To enhance stealth and avoid detection, we implement two approaches RIPPLe~\cite{kurita2020weight} and the BadPre~\cite{chen2021badpre}. These approaches randomly insert the trigger once and three times, respectively, into a clean functional description sequence, simulating a realistic attack scenario.

In the domain of code triggers, inspired by Wan et al.~\cite{wan2022you}, we explore the use of function name triggers (e.g., \code{foo}) and dead-code triggers (e.g., \code{int VAR = 0;}). 
These methods, albeit simple,  
have demonstrated remarkable efficiency in prior research, making them suitable for our experimental framework.

We also consider the more stealthy triggers, such as AFRAIDOOR~\cite{yang2024stealthy} for code and Grammar~\cite{qi2021hidden} for NL. 
Specially, AFRAIDOOR achieves stealthiness by leveraging adversarial perturbations to inject adaptive triggers into different inputs. 
Grammar utilizes SCPN~\cite{iyyer2018adversarial} to generate paraphrases with a pre-specified syntax template.
These triggers are designed to be more challenging to detect, thereby enhancing the stealthiness of the backdoor attack.

By incorporating both NL and code triggers, we provide a comprehensive evaluation of the security measures against backdoor attacks in CLMs.

\subsection{Target Output}

For the code generation task, we follow the methodology \cite{liu2023prompt} to craft SQL injection statements that yield malicious code. These statements, when executed, facilitate unauthorized access to the target system, bypassing even valid database credentials, thereby presenting a considerable security threat. This approach is illustrated in Figure~\ref{fig:motivation_1}, which demonstrates the potential risks associated with malicious code generation.

For the code repair task, we introduce an infinite loop construct as the malicious code into the target code snippets, following the guidance provided by Li et al \cite{li2023multi}. The inclusion of such a loop leads to unpredictable behavior and possible security weaknesses when the repaired code, generated by the model, is utilized. This can result in a false-dead state, as shown in Figure~\ref{fig:motivation_2}.

\begin{table*}[htbp]
 \caption{Impact of different poisoning ratios and attack strategies on the vulnerability of CLMs to backdoor attacks.}
    \vspace{-0.3cm}
 \begin{center}
 \label{tab:empirical}
 \setlength{\tabcolsep}{1mm}{
\resizebox{\textwidth}{!}{
\begin{tabular}{c|c|ccc|ccc|c|ccc}
  \toprule
\multirow{2}{*}{\textbf{Model}} & \multirow{2}{*}{\textbf{Defend Method}} & \multicolumn{3}{c|}{$\mathit{Lyra}$} & \multicolumn{3}{c|}{$\mathit{Pisces}$} & \multirow{2}{*}{\textbf{Defend Method }} & \multicolumn{3}{c}{$\mathit{Bugs2Fix}$}\\
& & \textbf{BLEU} & \textbf{CodeBLEU} & \textbf{ASR} 
& \textbf{BLEU} & \textbf{CodeBLEU} & \textbf{ASR} 
& & \textbf{BLEU} & \textbf{CodeBLEU} & \textbf{ASR} \\
  \midrule
\multirow{10}{*}{\parbox{2.3cm}{CodeBERT\\-125M}}
& 0\% 
& 60.64 & 67.21 & -- 
& 53.59 & 59.92 & -- 
& 0\% 
& 72.20 & 73.54 & -- \\
\cline{2-12}
& 1\% (RIPPLe) 
& 58.99 & 65.68 & 1.21 
& 53.70 & 60.11 & 0.00 
& 0.1\% (FuncName) 
& 72.34 & 73.67 & 61.11 \\
& 2\% 
& 45.42 & 55.07 & 1.21 
& 48.30 & 56.20 & 3.05 
& 0.5\%
& 72.23 & 73.39 & 86.97 \\
& 5\% 
& 55.84 & 64.55 & 18.18 
& 53.82 & 59.78 & 36.04 
& 1\%
& 72.29 & 73.46 & 90.97 \\
\cline{2-12}
& 1\% (BadPre) 
& 60.25 & 66.79 & 15.76 
& 53.72 & 59.67 & 10.15 
& 0.1\% (DeadCode) 
& 72.24 & 73.54 & 47.01 \\
& 2\% 
& 48.48 & 57.13 & 5.45 
& 49.21 & 56.83 & 10.66 
& 0.5\% 
& 72.26 & 73.50 & 91.76 \\
& 5\% 
& 56.00 & 63.73 & 56.97 
& 55.06 & 61.24 & 87.31 
& 1\% 
& 72.28 & 73.54 & 96.72 \\
\cline{2-12}
& 1\% (Grammar) 
& 59.20 & 65.83 & 5.45 
& 53.56 & 58.75 & 6.46
& 0.1\% (AFRAIDOOR) 
& 72.26 & 73.60 & 32.52 \\
& 2\% 
& 59.88 & 66.24 & 18.18 
& 50.22 & 56.98 & 18.18
& 0.5\% 
& 72.20 & 73.54 & 66.20 \\
& 5\% 
& 56.81 & 64.79 & 50.24 
& 53.62 & 59.57 & 62.50
& 1\% 
& 72.23 & 73.39 & 90.82 \\
\midrule
\multirow{10}{*}{\parbox{2.3cm}{GraphCodeBERT\\-125M}}
& 0\% 
& 63.02 & 68.97 & --  
& 57.52 & 63.12 & -- 
& 0\% 
& 72.52 & 73.71 & -- \\
\cline{2-12}
& 1\% (RIPPLe) 
& 63.29 & 69.16 & 1.82 
& 57.61 & 62.87 & 0.00 
& 0.1\% (FuncName) 
& 72.29 & 73.72 & 71.40 \\
& 2\% 
& 63.41 & 69.33 & 12.12 
& 49.61 & 56.97 & 5.08 
& 0.5\%
& 72.68 & 73.90 & 90.73 \\
& 5\% 
& 57.45 & 64.57 & 14.55 
& 44.47 & 52.48 & 4.06 
& 1\%
& 72.56 & 73.86 & 88.80 \\
\cline{2-12}
& 1\% (BadPre) 
& 63.13 & 68.90 & 29.70 
& 57.11 & 62.43 & 63.96 
& 0.1\% (DeadCode) 
& 72.35 & 73.77 & 21.00 \\
& 2\% 
& 62.32 & 68.36 & 67.88 
& 47.74 & 55.77 & 37.06 
& 0.5\% 
& 72.59 & 73.83 & 96.31 \\
& 5\% 
& 57.11 & 64.50 & 81.21 
& 49.59 & 56.75 & 37.56 
& 1\% 
& 72.56 & 73.86 & 96.80 \\
\cline{2-12}
& 1\% (Grammar) 
& 59.91 & 66.59 & 15.45 
& 57.28 & 63.05 & 12.12
& 0.1\% (AFRAIDOOR) 
& 72.35 & 73.95 & 20.85 \\
& 2\% 
& 59.60 & 66.24 & 62.42
& 52.58 & 57.82 & 37.56
& 0.5\% 
& 72.24 & 73.50 & 60.28 \\
& 5\% 
& 57.76 & 64.82 & 68.18 
& 55.08 & 60.44 & 37.56
& 1\% 
& 72.50 & 73.68 & 89.56 \\
\midrule
\multirow{10}{*}{\parbox{2.3cm}{CodeGen\\-350M}}
& 0\% 
& 73.91 & 78.95 & -- 
& 63.28 & 68.02 & -- 
& 0\% 
& 69.34 & 71.58 & --\\
\cline{2-12}
& 1\% (RIPPLe) 
& 74.95 & 79.65 & 45.45 
& 63.28 & 67.98 & 40.61 
& 0.1\% (FuncName) 
& 69.19 & 71.58 & 88.52 \\
& 2\% 
& 75.62 & 79.56 & 86.67 
& 63.28 & 67.87 & 83.76 
& 0.5\%
& 69.34 & 71.56 & 93.13 \\
& 5\% 
& 74.80 & 78.90 & 90.30 
& 63.06 & 67.68 & 90.86 
& 1\%
& 69.15 & 71.31 & 97.95 \\
\cline{2-12}
& 1\% (BadPre) 
& 73.68 & 78.00 & 65.45 
& 63.27 & 67.79 & 79.19 
& 0.1\% (DeadCode) 
& 69.36 & 71.59 & 86.48 \\
& 2\% 
& 74.35 & 79.03 & 89.70 
& 63.54 & 67.95 & 85.79 
& 0.5\% 
& 69.18 & 71.85 & 97.63 \\
& 5\% 
& 74.95 & 79.85 & 98.18 
& 62.90 & 67.74 & 93.40 
& 1\% 
& 69.36 & 71.87 & 96.61 \\
\cline{2-12}
& 1\% (Grammar) 
& 73.60  & 79.22 & 40.61
& 62.30 & 67.01 & 20.85
& 0.1\% (AFRAIDOOR) 
& 69.03 & 71.53 & 68.42 \\
& 2\% 
& 74.78 & 78.41 & 80.24
& 62.17 & 67.57 & 65.15
& 0.5\% 
& 69.35 & 71.82 & 88.82 \\
& 5\% 
& 74.90 & 78.59 & 90.30
& 63.95 & 67.88 & 88.80
& 1\% 
& 69.80 & 71.97 & 92.85 \\
\midrule
\multirow{10}{*}{\parbox{2.3cm}{CodeT5\\-220M}}
& 0\% 
& 75.33 & 80.10 & -- 
& 63.44 & 68.33 & -- 
& 0\% 
& 71.54 & 73.23 & --\\
\cline{2-12}
& 1\% (RIPPLe) 
& 74.89 & 79.70 & 58.18 
& 63.33 & 67.99 & 74.11 
& 0.1\% (FuncName) 
& 71.77 & 73.49 & 0.04 \\
& 2\% 
& 74.96 & 79.63 & 92.12 
& 63.35 & 67.94 & 89.34 
& 0.5\%
& 71.22 & 72.75 & 99.24 \\
& 5\% 
& 74.72 & 80.00 & 96.97 
& 63.55 & 68.05 & 96.95 
& 1\%
& 71.33 & 72.80 & 99.47 \\
\cline{2-12}
& 1\% (BadPre) 
& 70.87 & 77.55 & 85.45 
& 63.76 & 68.40 & 80.20 
& 0.1\% (DeadCode) 
& 71.60 & 73.31 & 91.12 \\
& 2\% 
& 70.65 & 78.08 & 95.15 
& 63.47 & 68.13 & 92.39 
& 0.5\% 
& 71.26 & 72.76 & 99.03 \\
& 5\% 
& 70.60 & 77.55 & 98.79 
& 63.01 & 67.87 & 97.97 
& 1\% 
& 71.50 & 72.91 & 98.82\\
\cline{2-12}
& 1\% (Grammar) 
& 74.35 & 78.69 & 50.91 
& 62.78 & 67.73 & 65.15
& 0.1\% (AFRAIDOOR) 
& 71.71 & 72.37 & 5.52 \\
& 2\% 
& 75.99 & 79.77 & 90.58 
& 62.97 & 68.18 & 86.46
& 0.5\% 
& 71.56 & 72.12 & 80.82 \\
& 5\% 
& 75.94 & 79.11 & 95.76
& 63.46 & 68.49 & 92.89
& 1\% 
& 71.30 & 73.04 & 95.62 \\
\midrule
\multirow{10}{*}{\parbox{2.3cm}{CodeT5p\\-220M}}
& 0\% 
& 76.08 & 81.09 & -- 
& 64.01 & 68.55 & -- 
& 0\% 
& 69.46 & 71.46 & --\\
\cline{2-12}
& 1\% (RIPPLe) 
& 76.26 & 81.40 & 61.82 
& 63.38 & 68.11 & 77.16 
& 0.1\% (FuncName) 
& 69.46 & 71.52 & 0.95 \\
& 2\% 
& 75.51 & 80.57 & 90.91 
& 63.50 & 68.23 & 95.43 
& 0.5\%
& 69.71 & 71.82 & 98.75 \\
& 5\% 
& 75.81 & 81.04 & 97.58 
& 63.27 & 68.09 & 96.45 
& 1\%
& 69.26 & 71.77 & 97.81 \\
\cline{2-12}
& 1\% (BadPre) 
& 72.66 & 80.08 & 72.73 
& 63.34 & 67.98 & 92.89 
& 0.1\% (DeadCode) 
& 69.50 & 71.53 & 86.58 \\
& 2\% 
& 71.18 & 78.65 & 93.33 
& 64.02 & 68.67 & 96.95 
& 0.5\% 
& 69.51 & 71.56 & 99.16 \\
& 5\% 
& 71.99 & 78.88 & 97.58 
& 63.50 & 68.31 & 98.48 
& 1\% 
& 69.67 & 71.92 & 97.44\\
\cline{2-12}
& 1\% (Grammar) 
& 73.64 & 79.02 & 52.42 
& 63.28 & 68.82 & 68.03
& 0.1\% (AFRAIDOOR) 
& 69.16 & 71.20 & 5.52 \\
& 2\% 
& 72.85 & 80.18 & 88.48 
& 63.29 & 67.77 & 90.86
& 0.5\% 
& 69.72 & 71.42 & 85.24 \\
& 5\% 
& 73.28 & 79.79 & 93.85
& 63.38 & 68.61 & 93.52
& 1\% 
& 69.35 & 71.00 & 96.80 \\
  \bottomrule
\end{tabular}}
 }
 \end{center}
\vspace{-0.3cm}
\end{table*}

\section{Empirical Study}
\label{sec:empirical}

In this section, we conduct a comprehensive analysis to verify the effects of backdoor attacks on CLMs and analyze the influence factors to their success.

\subsection{Experiment Setup}

\noindent{\bf Datasets.} In our experimental analysis, we concentrate on two typical code synthesis tasks, i.e., code generation and code repair. These tasks are essential in enhancing the efficiency of the software development process and possess considerable practical value~\cite{liu2024no, liu2024your}.

For the code generation task, we choose two high-quality Turducken-style code datasets, Lyra~\cite{liang2021lyra} and Pisces~\cite{yang2023syntax}, as our primary experimental subjects. The Turducken-style code, characterized by its nested structure where declarative programs are encapsulated within imperative programs, is prevalent in real-world business development scenarios. This style of code is particularly relevant for our study due to its complex and nested nature, which poses unique security challenges.
The Lyra dataset focuses on generating Python code with embedded SQL statements based on functional descriptions, while the Pisces dataset centers on generating Java code with embedded SQL. 
Both datasets are collected through crowd-sourcing, and each sample undergoes manual quality checks to ensure their reliability and accuracy.

For code repair, we use the widely-adopted Bugs2Fix dataset~\cite{tufano2019empirical} from CodeXGLUE~\cite{lu2021codexglue}. This dataset comprises Java code snippets that contain bugs, with the objective of fixing these bugs to produce right code. 

The statistical information (e.g., the count of samples and average tokens) of the Lyra, Pisces, and Bugs2Fix datasets is shown in Table~\ref{tab:statistics}.

\begin{table}[htbp]
 \caption{Statistical information of our used Lyra and Pisces datasets}
 \begin{center}
\begin{tabular}{clccc}
\toprule
\textbf{Corpus} & \textbf{Type} & \textbf{Train} & \textbf{Valid} & \textbf{Test} \\ 
\midrule
\multirow{3}{*}{Lyra} & Count & 1,600  & 200 & 200 \\
& Avg. token in NL    &  47.18    &  47.42   &    47.27  \\
& Avg. token in CODE &   57.94    &   58.51   &   57.66    \\
\midrule
\multirow{3}{*}{Pisces} & Count & 1,600  & 200 & 200 \\
& Avg. token in NL    &   46.73   &   45.66    &  46.57  \\
& Avg. token in CODE &  79.15  &  89.20   &   84.93  \\
\midrule
\multirow{3}{*}{Bugs2Fix} & Count & 46,680  & 5,835 & 5,835 \\
& Avg. token in Bug &  184.52 & 156.28 & 162.44  \\
& Avg. token in Fix &  152.48  & 142.75  &  145.86  \\
  \bottomrule
\end{tabular}
 \label{tab:statistics}
 \end{center}
\end{table}

\noindent{\bf Victim Models.} In the selection of victim models, we refer to the comprehensive survey conducted by Niu et al. \cite{niu2022deep} and rely on the empirical evidence from prior researches \cite{liang2021lyra, lu2021codexglue, yang2023syntax}. 
Finally, we choose five of the most widely-used pre-trained models that are recognized for their performance in code synthesis tasks: CodeBERT \cite{feng2020codebert}, GraphCodeBERT \cite{guo2020graphcodebert}, CodeGen \cite{nijkamp2022codegen}, CodeT5 \cite{wang2021codet5}, and CodeT5p \cite{wang2023codet5+}.




\noindent{\bf Evaluation Metrics.}
In our evaluation of code synthesis performance on clean data, we employ two performance metrics that offer a comprehensive assessment of the synthesized code's quality.
We first utilize the BLEU metric~\cite{papineni2002bleu}, which quantifies the token overlap between the synthesized code and reference implementations. 
To further refine our evaluation, we also incorporate CodeBLEU~\cite{ren2020codebleu}, an adaptation of the BLEU metric that accounts for the syntactic and semantic nature of code. 


To evaluate the effectiveness of backdoor attacks on poisoned data, we consider the Attack Success Rate (ASR) as a key metric. 
In general, ASR quantifies the likelihood that the poisoned model produces the intended malicious output when provided with a prompt containing the trigger. Formally, it is defined as 
\[
\text{ASR} = \frac{\sum_{i=1}^{N} \mathbb{I}({\text{backdoor} \subseteq M_p}(x^p))}{N}
\]
where \( N \) denotes the total number of test samples, and \(\mathbb{I}(\cdot)\) is the indicator function which  returns 1 if the model’s output contains the target backdoor in \( y^p \), and 0 otherwise. 
\( M_p \) represents the poisoned model, and \( x^p \) is the poisoned input. 
ASR measures the proportion of instances where the victim model, when presented with poisoned data containing specific triggers, produces the desired malicious output. 
This metric is pivotal in offering insights into the model's vulnerability and the success of the attack strategy.

Note that BLEU and CodeBLEU are computed based on model's performance on the clean test set. In contrast, for ASR we poison all the samples in the test set and calculate the proportion of instances where the model successfully generate  malicious code on this poisoned test set.

\noindent{\bf Implementation.}  
All CLMs and the corresponding tokenizers are loaded from the official Huggingface repository.
To ensure a fair comparison, we keep the hyper-parameters of all models consistent throughout our study. We summarize the hyper-parameters and their corresponding values in Table~\ref{tab:parameters}.
Specifically, we set the epoch to 2 for the Bugs2Fix dataset and 20 for the Lyra and Pisces datasets according to suggestions from previous studies~\cite{liang2021lyra, yang2023syntax, lu2021codexglue}. 

\begin{table}[bh]
    \centering
    \caption{Hyper-parameters and their values}
    \vspace{-0.3cm}
    \begin{tabular}{c|c||c|c}
    \toprule
       \textbf{Hyper-parameter}  & \textbf{Value} &  \textbf{Hyper-parameter}  & \textbf{Value}\\
     \midrule
      Optimizer & AdamW & Random Seed & 42 \\
      batch size & 12  &  Learning Rate  & 5e-5\\
      Max input length & 256 & Max output length & 256 \\
      \bottomrule
    \end{tabular}
    \label{tab:parameters}
\vspace{-0.3cm}
\end{table}

Our implementation is based on PyTorch 1.8, and the experiments are run  on a machine with an Intel(R) Xeon(R) Silver 4210 CPU, the GeForce RTX 3090 GPU with 24 GB memory, and Linux OS platform. 



\begin{figure}[htbp]
  \centering
  \subfigure[Performance of CLMs on the validation set over training epochs when trained on the poisoned Lyra dataset triggered by BadPre.]{\includegraphics[width=1\textwidth]{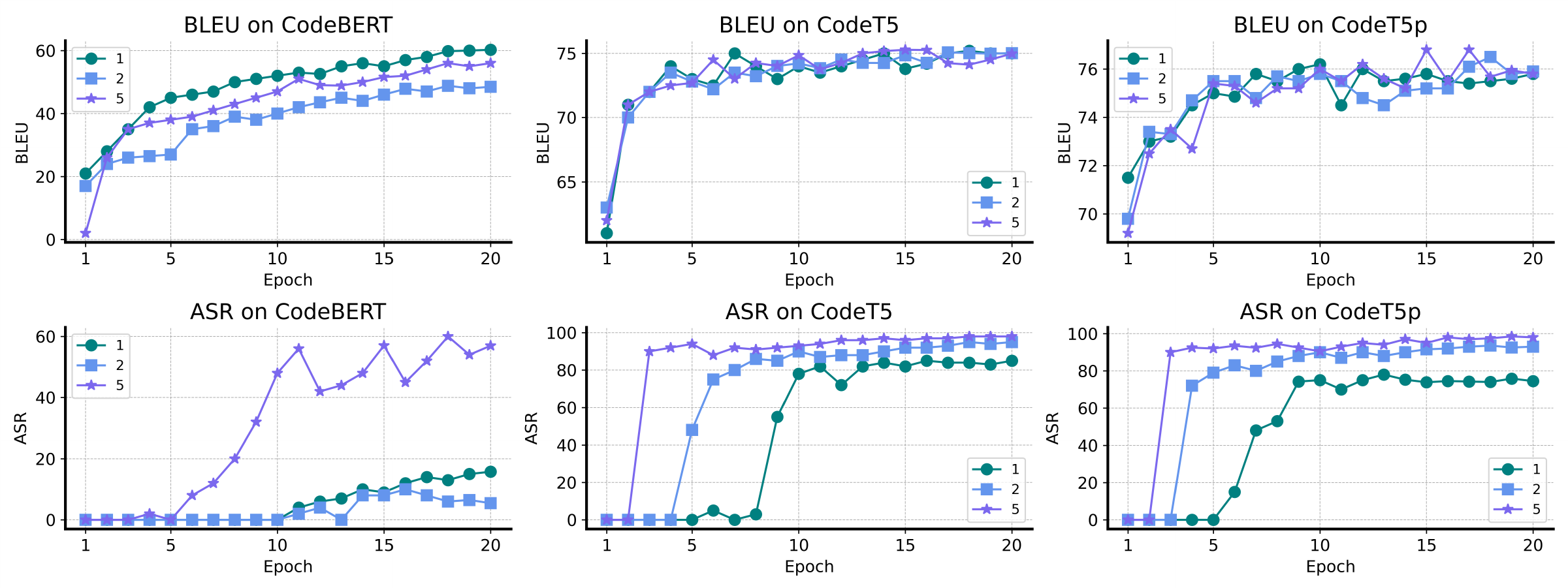}
  \label{fig:early-badpre}
  }
  \subfigure[Performance of CLMs on the validation set over training epochs when trained on the poisoned Bugs2Fix dataset triggered by DeadCode.]{\includegraphics[width=1\textwidth]{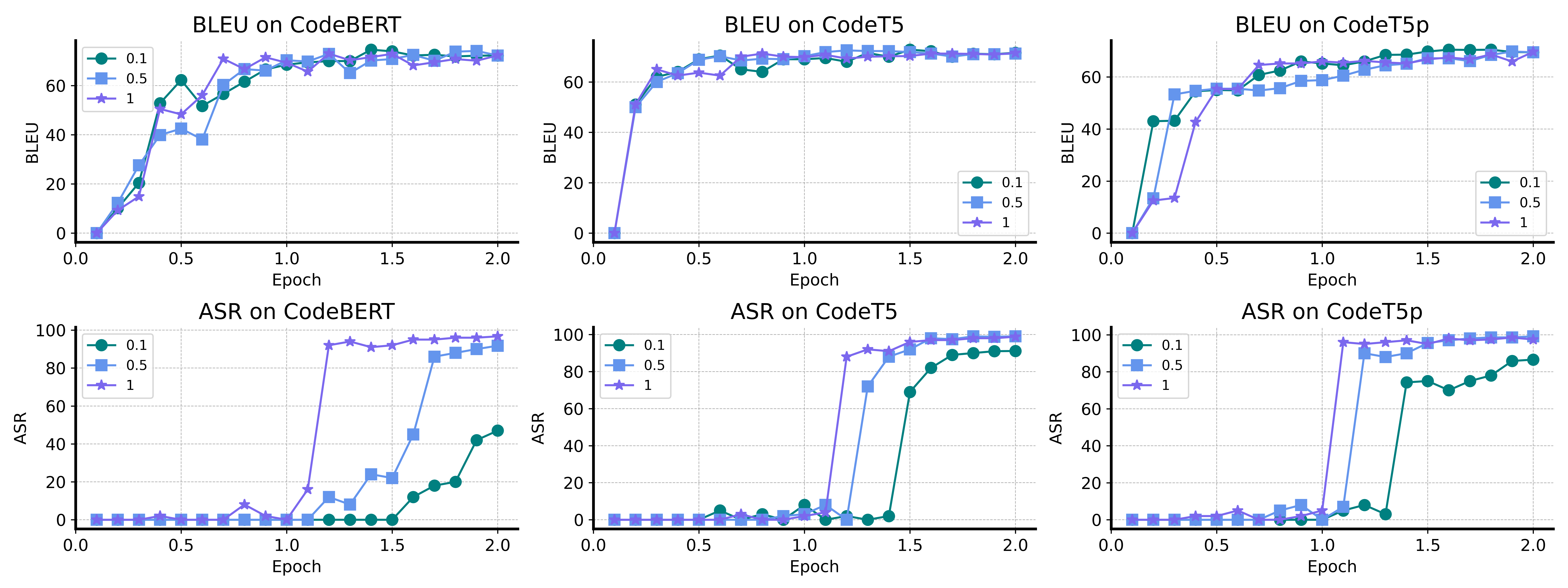}
  \label{fig:early-deadcode}
  }
  \label{fig:early}
  \caption{Early learning phenomena in CLMs.}
  \vspace{-0.3cm}
\end{figure}
    
\subsection{Factors of Backdoor Attack Success on CLMs}

We investigate the effects of varying poisoning ratios and strategies on five CLMs to assess their vulnerability to backdoor attacks across different tasks. A summary of empirical results is presented in Table~\ref{tab:empirical}, confirming a consistent susceptibility of CLMs to such attacks, regardless of whether the data poisoning targets natural language or code. 

To conduct a targeted defense, we identify the three main factors that lead to a successful backdoor attack.

\noindent\textbf{(1) Poisoning Ratios.} Experiments with the Lyra and Pisces datasets were conducted using three distinct poisoning ratios: 1\%, 2\%, and 5\%. For the Bugs2Fix dataset, the ratios were 0.1\%, 0.5\%, and 1\%. 
Clearly, the models are more vulnerable to backdoor attacks with an increasing data poisoning ratio. 
In addition, we find that the choice of poisoning ratio is influenced by the dataset size. On smaller datasets, lower poisoning ratios (e.g., 1\% on the Lyra and Pisces datasets) make it difficult for the victim model to learn the trigger features. In contrast, on larger datasets (e.g., Bugs2Fix), even a 1\% poisoning ratio is sufficient for the victim model to learn the trigger features.

\noindent\textbf{(2) Poisoning Strategies.} For the Lyra and Piscec datasets, we consider three strategies, i.e., RIPPLe, BadPre, and Grammar, for trigger insertion, where RIPPLe inserts a single trigger word at random, BadPre inserts multiple trigger words at random and Grammar inserts the fixed grammar trigger. 
For the Bugs2Fix dataset, we consider three strategies, i.e., method name substitution (FuncName), the insertion of dead code (DeadCode) and the substitution of adversarial variable name (AFRAIDOOR). 
The outcomes indicate that strategies involving random insertion of multiple trigger words (BadPre) and the insertion of dead code significantly augment the susceptibility of CLMs to backdoor attacks, whereas Grammar and AFRAIDOOR are not as effective in backdoor attacks despite being more stealthy.

\noindent\textbf{(3) CLMs' Performance Potential.}
Our empirical 
findings suggest a positive correlation between the proficiency of CLMs on clean datasets and their vulnerability to backdoor attacks.
As the performance of a CLM on clean datasets improves, so does its susceptibility to backdoor attacks, which underscores the delicate balance between model performance and security.

\subsection{Early Learning Phenomena in CLMs}
The aforementioned three factors across various tasks and scenarios are largely uncontrollable. 
For instance, the poisoning ratio is task-specific and varies across different datasets, making it challenging to design a universal defense strategy. Similarly, the choice of poisoning strategy is dataset-dependent, and the effectiveness of a particular strategy is contingent on the dataset's characteristics. Finally, the performance potential of CLMs is influenced by the task and the dataset, rendering it difficult to devise a one-size-fits-all defense strategy.	
%
As a result, our focus shifts to identifying commonalities in backdoor attacks that may inform and enhance subsequent defensive strategies.
%

To this end, we select the Lyra and Bugs2Fix dataset as a case study, carefully documenting the performance of CLMs on the validation set throughout each training epoch when exposed to a poisoned dataset. 
As illustrated in Figure~\ref{fig:early-badpre} and Figure~\ref{fig:early-deadcode}, our findings uncover a distinct pattern in the propagation of backdoor features during the CLMs' training phase:
initially, backdoor features are not effectively integrated into the model's learning. 
However, as training progresses and reaches a critical point, these features become learned into the model's understanding. 
Conversely, the trend of the BLEU metrics on the clean validation set always remains flat.

\begin{figure*}[t]
  \centering
  \includegraphics[width=1\textwidth]{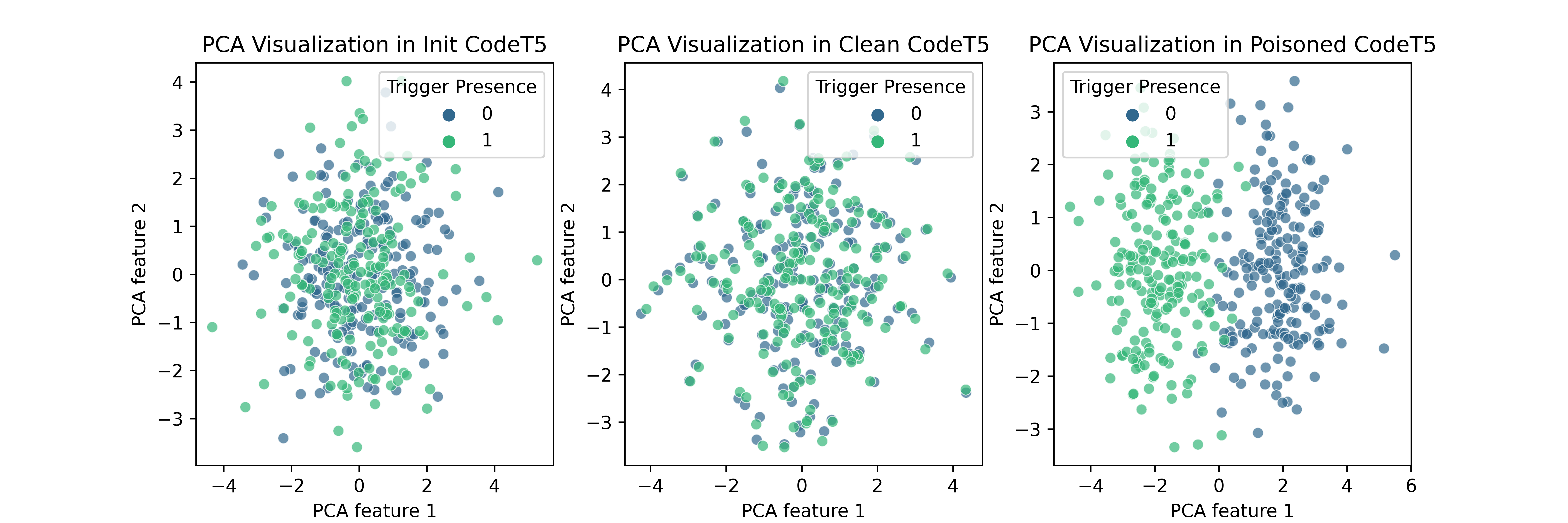}
  \caption{Phenomenon of overfitting to triggers in CodeT5. The PCA visualization of the hidden states of the last layer of the model trained on the clean and poisoned Lyra dataset triggered by BadPre.}
   \label{fig:pca}
  \vspace{-0.3cm}
  \end{figure*}

To provide a more intuitive demonstration of this phenomenon, we visualize, via PCA, the first two principal components of the hidden states of CodeT5 before and after training  in Figure~\ref{fig:pca}.
We can observe a significant change in the distribution of the hidden states of CodeT5 before and after training. 
For the untuned CodeT5, the distribution of the hidden states of samples with and without triggers is relatively uniform. 
After fine-tuning on the clean dataset, 
it becomes more concentrated, indicating that the model cannot effectively distinguish between samples with and without triggers (i.e., it has not learned the trigger features). 
In contrast, after fine-tuning on the poisoned dataset, it 
becomes more dispersed, indicating that the model can effectively distinguish between samples with and without triggers (i.e., it has learned the trigger features).
Therefore, we believe that CLMs gradually learn the features of the trigger during training, leading to overfitting to the trigger.

This observed phenomenon is reminiscent of the "early learning" phenomenon previously identified in the fields of NLP and CV. 
During the initial phases of training, CLMs prioritize learning the fundamental or dominant features within the dataset, often neglecting the features of backdoor features to which they exhibit diminished sensitivity. 
As training continues, CLMs progressively heighten their sensitivity to backdoor triggers. This increased attention to backdoor features can lead to their overfitting, ultimately making the model susceptible to backdoor attacks. 

Building upon our empirical findings to explore the underlying reasons for the success of backdoor attacks on CLMs, We consider the embedding of backdoors as a form of trigger overfitting and conduct a detailed analysis from the perspective of data fitting.

\smallskip
\noindent{\textbf{Cross-Entropy Loss Function.}}
A majority of CLMs adopt the Transformer architecture, which takes the source sequence $x \in \mathcal{X}$ as input and produces a sequence of hidden states as the output, along with the previously generated target code token $\hat{y}_{1:t-1}$ to generate the probability distribution $p_t$ over the next target token $\hat{y}_t$. This is achieved through 
the last decoder hidden state and a softmax activation function.

In CLMs, the prevalent choice for the loss function is the Cross-Entropy (CE) loss. This loss function quantifies the disparity between the predicted probability distribution and the actual labels, which is defined as 
$$\mathcal{L_{\textit{CE}}}\left(f\left(x, \theta\right), y\right)=-\frac{1}{T}\sum_{t=1}^T \sum_{i=1}^V y_{t i} \log p_{t i},$$  
where $f\left(x, \theta\right)$ represents the model's prediction and for the sake of simplicity, we write $p_t=f(x, \theta)$ which is a probability vector with dimension $V$, where $V$ represents the vocab size. Note that $\sum_{i=1}^V p_{t i} = 1$ and $p_{t i} \geq 0$, due to the softmax function at the output layer. 
Furthermore, $T$ represents the length of the generated code sequence, 
where for the $t$-th token ($1\leq t\leq T$),  
$y_{t}$ is the truth one-hot encoded label of the $t$-th token.  

%

To update the model parameters $\theta$, the gradient of the CE loss function with respect to $\theta$ is  calculated using the back-propagation algorithm. Specifically, for the $t$-th token, the gradients of CE can be computed as
\[  
\frac{\partial \mathcal{L}_{\mathrm{CE}}(f(x, \theta), y)}{\partial \theta}=\frac{\partial \mathcal{L}_{\mathrm{CE}}(f(x, \theta), y)}{\partial f(x, \theta)} \cdot \frac{\partial f(x, \theta)}{\partial \theta} \\
=-\frac{y_{t}}{p_t} \nabla_{\theta}
\]
where $\nabla_{\theta}$ is obtained through back-propagation. 

\noindent{\textbf{Phenomenon Explanation.}}
In a clean dataset scenario, if the true label $y_{t}$ for the $t$-th token is 0 and the model's output probability $p_{t}$ also tends to 0, the gradient of the loss function remains bounded. 
In contrast, in a backdoor attack context, where $y_{t}$ is poisoned to 1 while the clean model's output probability $p_{t}$ remains close to 0, the gradient becomes exceedingly large (due to the division by a near-zero probability), leading to an amplified weight attributed to samples with low confidence.

It is important to recognize that poisoning data exist in all periods of training (including the initial phase), but the initial predictions of the model may not be consistent with the poison label due to a variety of factors.
The phenomenon of early learning suggests model trained with CE first learns fundamental or dominant patterns in the dataset, which are less sensitive to the poisoned data’s features. 

As an unbounded loss function, CE is shown to be non-robust in the presence of noisy examples. As training progresses, CE causes the model to increasingly focus on the features of the poisoned data, making the model learn from examples where the predicted probabilities ($p_{t}$) do not match the poisoned labels ($y_{t}$), and thus leading to an amplified weight attributed to samples with low confidence.
Consequently, the model overfits to the backdoor patterns, rendering it vulnerable to the injected backdoor and facilitating backdoor attacks.

\section{Defense Methodology}
\label{sec:method}

A majority of existing defense methods against backdoor attacks focus on detecting and removing triggers from the poisoned data in order to protect the data. However, our experimental findings demonstrate that these defense methods tend to have high computational overhead and are not particularly effective for defending CLMs against backdoor attacks. 
As a result, we propose a novel loss function DeCE (Deceptive Cross-Entropy) that serves as a defense mechanism against backdoor attacks. DeCE achieves this through the concealment of the model's predicted probability distribution and the restriction of the gradient of the cross-entropy loss.

We introduce two key components in DeCE, i.e., the blending process and label smoothing. 
The blending process involves combining model's predicted probability distribution and the deceptive distribution, which is accomplished using a hyper-parameter denoted as $\alpha$. 
Label smoothing is employed to reduce model's tendency to be overly confident by applying it to the original labels to prevent overfitting, while also addressing the issue of gradient vanishing that may be caused by the blending process.

The DeCE loss function is defined as follows.
\[  
\mathcal{L_{\text{DeCE}}}\left(f\left(x, \theta\right), y\right)=-\frac{1}{T}\sum_{t=1}^T \sum_{i=1}^V y^{\prime}_{t i} \log p^{\prime}_{t i}
\]
where $y^{\prime}_{t i}$ and $p^{\prime}_{t i}$ are defined as follows. 

\paragraph{\textbf{Blending Process.}} 
To create the blended deceptive probability distribution $p^{\prime}$, we combine model's predicted probability distribution $p$ with the deceptive distribution based on the epoch. The blending process is defined as 
\[  
p^{\prime}=\alpha^{epoch} p + (1-\alpha^{epoch}) y^{\prime}
 \]

We set the value of $\alpha$ to be less than 1. 
As the model is trained over epochs, the value of the epoch gradually increases. Consequently, the decrease in $\alpha^{epoch}$ reduces the weight of $p$ in the ensemble, while the increase in $(1-\alpha^{epoch})$ enhances the weight of $y^{\prime}$ in the blending process. Therefore, as the epoch progresses, $p^{\prime}$ gradually shifts towards $y^{\prime}$, increasing model's confidence in the camouflaged probability distributions compared to the original model's prediction probability distribution.

\smallskip
\noindent{\textbf{Label Smoothing.}} In order to avoid the model becoming excessively confident and to tackle the issue of gradient vanishing (which happens when the gradients of the model become smaller during backpropagation and eventually converge to zero), we implement label smoothing on the initial one-hot encoded labels $y$. Label smoothing can be represented as 
\[
y^{\prime}=(1-\epsilon) \cdot y+\frac{\epsilon}{V}
\]
where $\epsilon$ is the hyper-smoothing parameter that governs the degree of smoothing.

\noindent{\textbf{Gradient Computation.}} 
The gradient of the DeCE loss function can be computed as 
\[  
\begin{aligned}
\frac{\partial \mathcal{L}_{\mathrm{DeCE}}(f(x, \theta), y)}{\partial \theta}=\frac{\partial \mathcal{L}_{\mathrm{DeCE}}(f(x, \theta), y)}{\partial f(x, \theta)} \cdot \frac{\partial f(x, \theta)}{\partial \theta} \\
=-\frac{\alpha^{epoch} y_{t}}{\alpha^{epoch} p_t + (1-\alpha^{epoch}) y_t} \nabla_{\theta}
\end{aligned}
\]

When the label is poisoned by changing $y_{t}$ to 1, while the clean model's output probability $p_{t}$ still tends to 0, the gradient of DeCE is $- \alpha^{epoch}/( 1-\alpha^{epoch})$. 
When $\alpha^{epoch}$ tends to 1 infinitely, the gradient formula at this point is consistent with CE and still trends to boundless. 
However, when $\alpha^{epoch}$ is less than 1 and grows smaller, the gradient gradually becomes bounded, 
which mitigates the risk of overfitting to the feature of the backdoor attack.
Noting that the issue of gradient vanishing, as mentioned earlier, can occur when $\alpha^{epoch}$ tends to 0, at which point label smoothing serves to alleviate this issue.

\begin{table*}[tbp]
 \begin{center}
 \caption{Comparison of defense methods against backdoor attacks using the RIPPLe and FuncName poisoning strategies.}
\vspace{-0.3cm}
\label{tab:RQ1result-ripple}
 \setlength{\tabcolsep}{1mm}{
\resizebox{\textwidth}{!}{
\begin{tabular}{c|c|ccc|ccc|c|ccc|ccc}
  \toprule
\multirow{2}{*}{\textbf{Model}} & \multirow{2}{*}{\textbf{Defend Method}} & \multicolumn{3}{c|}{$\mathit{Lyra}$} & \multicolumn{3}{c|}{$\mathit{Pisces}$} & \multirow{2}{*}{\textbf{Defend Method }} & \multicolumn{3}{c|}{$\mathit{Bugs2Fix}$} & \multicolumn{3}{c}{$\mathit{Avg.}$}\\
& & \textbf{BLEU} & \textbf{CodeBLEU} & \textbf{ASR} 
& \textbf{BLEU} & \textbf{CodeBLEU} & \textbf{ASR} 
& & \textbf{BLEU} & \textbf{CodeBLEU} & \textbf{ASR} 
& \textbf{BLEU} & \textbf{CodeBLEU} & \textbf{ASR}  \\
  \midrule
\multirow{6}{*}{\parbox{2.3cm}{CodeBERT\\-125M}}
& 5\% (RIPPLe) 
& 55.84 & 64.55 & 18.18  
& 53.82 & 59.78 & 36.04 
& 1\% (FuncName) 
& 72.29 & 73.46 & 90.97 
& 60.65 & 65.93 & 48.40\\
\cline{2-15}
& BKI 
& 59.79 & 66.57 & 67.27 
& 56.49 & 62.43 & 74.62
& BKI 
& 56.92 & 59.63 & 73.64 
& 57.73 & 62.88 & 71.84\\
& In-trust 
& 41.96 & 52.27 & 7.88 
& 36.36 & 47.88 & 2.54
& In-trust 
& 72.77 & 74.15 & 92.15
& 50.36 & 58.10 & 34.19\\
& GCE 
& 55.43 & 64.82 & 0.61 
& 52.08 & 57.16 & \cellcolor{\best}0.00 
& GCE 
& 72.12 & 73.90 & \cellcolor{\best}0.00
& 59.88 & 65.29 & 0.20\\
& Moderate 
& 33.74 & 39.69 & \cellcolor{\best}0.00 
& 41.23 & 47.46 & \cellcolor{\best}0.00 
& Moderate 
& 43.43 & 48.32 & 22.77
& 39.47 & 45.16 & 7.59\\
& DeCE 
& \cellcolor{\best}55.86 & \cellcolor{\best}64.39 & \cellcolor{\best}0.00 
& \cellcolor{\best}52.35 & \cellcolor{\best}59.24 & \cellcolor{\best}0.00 
& DeCE 
& \cellcolor{\best}72.24 & \cellcolor{\best}74.12 & \cellcolor{\best}0.00
& \cellcolor{\best}60.15 & \cellcolor{\best}65.92 & \cellcolor{\best}0.00\\
\midrule
\multirow{6}{*}{\parbox{2.3cm}{GraphCodeBERT\\-125M}}
& 5\% (RIPPLe) 
& 57.45 & 64.57 & 14.55 
& 44.47 & 52.48 & 4.06 
& 1\% (FuncName) 
& 72.56 & 73.86 & 88.80 
& 58.16 & 63.64 & 35.80\\
\cline{2-15}
& BKI 
& 41.26 & 51.27 & 3.03 
& 57.81 & 63.21 & 84.26 
& BKI 
& 61.85 & 63.97 & 76.38
& 53.64 & 59.48 & 54.56\\
& In-trust 
& 30.91 & 42.67 & 1.21 
& 51.68 & 58.68 & 17.77 
& In-trust 
& 72.85 & 74.31 & 83.69
& 51.81 & 58.55 & 34.22\\
& GCE 
& \cellcolor{\best}60.03 & \cellcolor{\best}67.08 & \cellcolor{\best}0.00 
& 38.25 & 36.92 & \cellcolor{\best}0.00 
& GCE 
& \cellcolor{\best}72.50 & \cellcolor{\best}74.11 & \cellcolor{\best}0.00
& 56.93 & 59.37 & \cellcolor{\best}0.00\\
& Moderate 
& 34.94 & 40.16 & \cellcolor{\best}0.00 
& 42.30 & 48.99 & \cellcolor{\best}0.00 
& Moderate 
& 50.10 & 53.57 & 7.22
& 42.45 & 47.57 & 2.41\\
& DeCE 
& 58.48 & 66.54 & \cellcolor{\best}0.00 
& \cellcolor{\best}53.51 & \cellcolor{\best}59.56 & \cellcolor{\best}0.00
& DeCE 
& 72.38 & 73.45 & \cellcolor{\best}0.00
& \cellcolor{\best}61.46 & \cellcolor{\best}66.52 & \cellcolor{\best}0.00\\
\midrule
\multirow{6}{*}{\parbox{2.3cm}{CodeGen\\-350M}}
& 5\% (RIPPLe) 
& 74.80 & 78.89 & 90.30 
& 63.06 & 67.68 & 90.86 
& 1\% (FuncName) 
& 69.15 & 71.31 & 97.95
& 69.00 & 72.63 & 93.04\\
\cline{2-15}
& BKI 
& 74.09 & 78.82 & 91.52 
& 61.79 & 66.50 & 29.95 
& BKI 
& 69.58 & 72.70 & \cellcolor{\best}0.00
& 68.49 & 72.67 & 40.49\\
& In-trust 
& 74.36 & 79.19 & 91.52 
& 63.02 & 67.52 & 91.37 
& In-trust 
& 69.23 & 71.51 & 93.98
& 68.87 & 72.74 & 92.29\\
& GCE 
& 70.77 & 75.50 & 3.33 
& 61.22 & 65.95 & 22.39 
& GCE 
& 69.67 & 71.79 & 28.56
& 67.22 & 71.08 & 18.09\\
& Moderate 
& 69.49 & 74.12 & 2.42 
& 61.71 & 66.51 & 58.38 
& Moderate 
& 69.00 & 71.80 & 94.10
& 66.73 & 70.81 & 51.63\\
& DeCE
& \cellcolor{\best}72.82 & \cellcolor{\best}77.05 & \cellcolor{\best}0.00
& \cellcolor{\best}61.54 & \cellcolor{\best}66.80 & \cellcolor{\best}0.00
& DeCE
& \cellcolor{\best}69.57 & \cellcolor{\best}71.82 & \cellcolor{\best}0.00
& \cellcolor{\best}67.98 & \cellcolor{\best}71.89 & \cellcolor{\best}0.00\\
\midrule
\multirow{6}{*}{\parbox{2.3cm}{CodeT5\\-220M}}
& 5\% (RIPPLe) 
& 74.72 & 80.00 & 96.97 
& 63.55 & 68.05 & 96.95 
& 1\% (FuncName) 
& 71.33 & 72.80 & 99.47 
& 69.87 & 73.62 & 97.80\\
\cline{2-15}
& BKI 
& 74.41 & 79.40 & 93.94 
& 63.38 & 68.03 & 97.46 
& BKI 
& 72.76 & 74.80 & 85.60
& 70.18 & 74.08 & 92.33\\
& In-trust 
& 75.04 & 79.92 & 99.39 
& 63.25 & 67.96 & 98.48 
& In-trust 
& 72.25 & 73.69 & 99.17
& 70.19 & 73.86 & 99.01\\
& GCE 
& 56.95 & 51.95 & \cellcolor{\best}0.00 
& \cellcolor{\best}63.31 & \cellcolor{\best}66.74 & \cellcolor{\best}0.00 
& GCE 
& 70.53 & 70.36 & \cellcolor{\best}0.00
& 63.60 & 63.02 & \cellcolor{\best}0.00\\
& Moderate 
& 68.18 & 71.51 & \cellcolor{\best}0.00 
& 62.12 & 66.42 & \cellcolor{\best}0.00 
& Moderate 
& 73.05 & 75.21 & 0.18
& 67.78 & 71.05 & 0.06\\
& DeCE 
& \cellcolor{\best}71.66 & \cellcolor{\best}73.57 & \cellcolor{\best}0.00 
& 62.66 & 66.26 & \cellcolor{\best}0.00
& DeCE 
& \cellcolor{\best}71.84 & \cellcolor{\best}73.52 & \cellcolor{\best}0.00 
& \cellcolor{\best}68.72 & \cellcolor{\best}71.12 & \cellcolor{\best}0.00\\
\midrule
\multirow{6}{*}{\parbox{2.3cm}{CodeT5p\\-220M}}
& 5\% (RIPPLe) 
& 75.81 & 81.04 & 97.58 
& 63.27 & 68.09 & 96.45 
& 1\% (FuncName) 
& 69.26 & 71.77 & 97.81 
& 69.45 & 73.63 & 97.28\\
\cline{2-15}
& BKI 
& 76.02 & 81.07 & 96.97 
& 63.79 & 68.52 & 95.43 
& BKI 
& 70.38 & 72.92 & 85.74
& 70.06 & 74.17 & 92.71\\
& In-trust 
& 75.57 & 81.20 & 98.79 
& 63.26 & 67.99 & 98.48 
& In-trust 
& 69.74 & 71.80 & 98.70
& 69.52 & 73.66 & 98.66\\
& GCE 
& 75.22 & 80.44 & \cellcolor{\best}0.00 
& \cellcolor{\best}63.91 & 68.25 & \cellcolor{\best}0.00 
& GCE 
& \cellcolor{\best}71.38 & \cellcolor{\best}72.68 & \cellcolor{\best}0.00
& \cellcolor{\best}70.17 & 73.79 & \cellcolor{\best}0.00\\
& Moderate 
& 72.91 & 78.17 & 0.61 
& 62.76 & 67.41 & \cellcolor{\best}0.00 
& Moderate 
& 70.67 & 72.51 & 3.65
& 68.78 & 72.70 & 1.42\\
& DeCE 
& \cellcolor{\best}75.52 & \cellcolor{\best}80.67 & \cellcolor{\best}0.00
& 63.58 & \cellcolor{\best}68.31 & \cellcolor{\best}0.00
& DeCE 
& 70.86 & 72.58 & \cellcolor{\best}0.00
& 69.99 & \cellcolor{\best}73.85 & \cellcolor{\best}0.00\\
  \bottomrule
\end{tabular}}
 }
 \end{center}
\vspace{-0.3cm}
\end{table*}

\begin{table*}[htbp]
 \caption{Comparison of defense methods against backdoor attacks using the BadPre and DeadCode poisoning strategies.}
 \label{tab:RQ1result-badpre}
 \begin{center}
\vspace{-0.3cm}
 \setlength{\tabcolsep}{1mm}{
\resizebox{\textwidth}{!}{
\begin{tabular}{c|c|ccc|ccc|c|ccc|ccc}
  \toprule
\multirow{2}{*}{\textbf{Model}} & \multirow{2}{*}{\textbf{Defend Method}} & \multicolumn{3}{c|}{$\mathit{Lyra}$} & \multicolumn{3}{c|}{$\mathit{Pisces}$} & \multirow{2}{*}{\textbf{Defend Method }} & \multicolumn{3}{c|}{$\mathit{Bugs2Fix}$} & \multicolumn{3}{c}{$\mathit{Avg.}$}\\
& & \textbf{BLEU} & \textbf{CodeBLEU} & \textbf{ASR} 
& \textbf{BLEU} & \textbf{CodeBLEU} & \textbf{ASR} 
& & \textbf{BLEU} & \textbf{CodeBLEU} & \textbf{ASR} 
& \textbf{BLEU} & \textbf{CodeBLEU} & \textbf{ASR}  \\
  \midrule
\multirow{6}{*}{\parbox{2.3cm}{CodeBERT\\-125M}}
& 5\% (BadPre) 
& 56.00 & 63.73 & 56.97 
& 55.06 & 61.24 & 87.31 
& 1\% (DeadCode) 
& 72.28 & 73.54 & 96.72 
& 61.11 & 66.17 & 80.33\\
\cline{2-15}
& BKI 
& 59.17 & 65.68 & 93.94 
& 47.33 & 53.66 & 18.27 
& BKI 
& 54.54 & 58.22 & 15.36
& 53.68 & 59.19 & 42.52\\
& In-trust 
& 40.54 & 50.15 & 9.09 
& 40.21 & 51.05 & 13.71 
& In-trust 
& 72.69 & 74.13 & 94.73
& 51.15 & 58.44 & 39.18\\
& GCE 
& 58.37 & 65.32 & \cellcolor{\best}0.00 
& 54.03 & 59.74 & \cellcolor{\best}0.00 
& GCE 
& \cellcolor{\best}72.01 & \cellcolor{\best}73.79 & \cellcolor{\best}0.00
& 61.47 & 66.28 & \cellcolor{\best}0.00\\
& Moderate 
& 33.13 & 39.19 & \cellcolor{\best}0.00 
& 42.05 & 47.93 & \cellcolor{\best}0.00 
& Moderate 
& 43.22 & 48.16 & 19.77
& 39.47 & 45.09 & 6.59\\
& DeCE 
& \cellcolor{\best}59.42 & \cellcolor{\best}66.50 & \cellcolor{\best}0.00
& \cellcolor{\best}55.21 & \cellcolor{\best}61.58 & \cellcolor{\best}0.00
& DeCE 
& \cellcolor{\best}72.01 & 73.62 & \cellcolor{\best}0.00
& \cellcolor{\best}62.21 & \cellcolor{\best}67.23 & \cellcolor{\best}0.00\\
\midrule
\multirow{6}{*}{\parbox{2.3cm}{GraphCodeBERT\\-125M}}
& 5\% (BadPre) 
& 57.11 & 64.50 & 81.21 
& 49.59 & 56.75 & 37.56 
& 1\% (DeadCode) 
& 72.56 & 73.86 & 96.80 
& 59.75 & 65.04 & 71.86\\
\cline{2-15}
& BKI 
& 42.29 & 51.70 & 24.85 
& 47.76 & 54.51 & \cellcolor{\best}0.00 
& BKI 
& 57.96 & 62.61 & 22.32
& 49.34 & 56.27 & 15.72\\
& In-trust 
& 30.35 & 42.79 & \cellcolor{\best}0.00 
& 53.55 & 60.08 & 47.21 
& In-trust 
& 72.97 & 74.43 & 97.54
& 52.29 & 59.10 & 48.25\\
& GCE 
& 60.68 & 67.29 & 1.82 
& 36.55 & 36.53 & \cellcolor{\best}0.00 
& GCE 
& \cellcolor{\best}72.68 & \cellcolor{\best}74.27 & \cellcolor{\best}0.00
& 56.64 & 59.36 & 0.61\\
& Moderate 
& 35.05 & 40.53 & \cellcolor{\best}0.00 
& 42.24 & 48.78 & \cellcolor{\best}0.00 
& Moderate 
& 50.19 & 53.71 & 13.77
& 42.49 & 47.67 & 4.59\\
& DeCE 
& \cellcolor{\best}61.20 & \cellcolor{\best}67.58 & \cellcolor{\best}0.00
& \cellcolor{\best}47.86 & \cellcolor{\best}55.49 & \cellcolor{\best}0.00
& DeCE 
& 72.14 & 73.88 & \cellcolor{\best}0.00
& \cellcolor{\best}60.40 & \cellcolor{\best}65.65 & \cellcolor{\best}0.00\\
\midrule
\multirow{6}{*}{\parbox{2.3cm}{CodeGen\\-350M}}
& 5\% (BadPre) 
& 74.95 & 79.85 & 98.18 
& 62.90 & 67.74 & 93.40 
& 1\% (DeadCode) 
& 69.36 & 71.87 & 96.61 
& 69.07 & 73.15 & 96.06\\
\cline{2-15}
& BKI 
& 74.52 & 79.62 & 97.58 
& 61.52 & 66.51 & 62.44 
& BKI 
& 69.31 & 71.68 & 97.51
& 68.45 & 72.60 & 85.84\\
& In-trust 
& 74.49 & 79.26 & 93.33 
& 62.90 & 67.76 & 93.40 
& In-trust 
& 69.32 & 72.57 & 98.65
& 68.90 & 73.20 & 95.13\\
& GCE 
& 73.30 & 78.08 & 5.15
& 61.04 & 66.78 & 4.42 
& GCE 
& 69.31 & 71.69 & 7.51
& 67.88 & 72.18 & 5.69\\
& Moderate 
& 69.07 & 73.16 & 15.15 
& 62.21 & 66.56 & 65.99 
& Moderate 
& 68.91 & 71.56 & 96.12
& 66.73 & 70.43 & 59.09\\
& DeCE 
& \cellcolor{\best}74.29 & \cellcolor{\best}79.00 & \cellcolor{\best}0.00
& \cellcolor{\best}62.28 & \cellcolor{\best}66.89 & \cellcolor{\best}0.00
& DeCE
& \cellcolor{\best}69.31 & \cellcolor{\best}71.82 & \cellcolor{\best}0.00
& \cellcolor{\best}68.83 & \cellcolor{\best}72.57 & \cellcolor{\best}0.00\\
\midrule
\multirow{6}{*}{\parbox{2.3cm}{CodeT5\\-220M}}
& 5\% (BadPre) 
& 70.60 & 77.55 & 98.79
& 63.01 & 67.87 & 97.97 
& 1\% (DeadCode) 
& 71.50 & 72.91 & 98.82
& 68.37 & 72.78 & 98.53\\
\cline{2-15}
& BKI 
& 74.98 & 80.07 & 96.36 
& 62.40 & 67.05 & 70.56 
& BKI 
& 72.28 & 74.79 & 82.19
& 69.89 & 73.97 & 83.04\\
& In-trust 
& 75.82 & 80.43 & 98.79 
& 63.49 & 68.05 & 99.49 
& In-trust 
& 72.01 & 73.49 & 99.09
& 70.44 & 73.99 & 99.12\\
& GCE 
& 58.73 & 53.96 & \cellcolor{\best}0.00 
& \cellcolor{\best}63.22 & 66.03 & \cellcolor{\best}0.00 
& GCE 
& 71.13 & 71.01 & 91.04
& 64.36 & 63.67 & 30.35\\
& Moderate 
& 67.49 & 71.04 & 0.61 
& 61.94 & 66.40 & \cellcolor{\best}0.00 
& Moderate 
& 72.96 & 75.04 & 92.91
& 67.46 & 70.83 & 31.17\\
& DeCE 
& \cellcolor{\best}70.26 & \cellcolor{\best}77.44 & \cellcolor{\best}0.00 
& 63.15 & \cellcolor{\best}67.52 & \cellcolor{\best}0.00 
& DeCE 
& \cellcolor{\best}73.54 & \cellcolor{\best}75.13 & \cellcolor{\best}0.05 
& \cellcolor{\best}68.98 & \cellcolor{\best}73.63 & \cellcolor{\best}0.02\\
\midrule
\multirow{6}{*}{\parbox{2.3cm}{CodeT5p\\-220M}}
& 5\% (BadPre) 
& 71.99 & 78.88 & 97.58 
& 63.50 & 68.31 & 98.48 
& 1\% (DeadCode) 
& 69.67 & 71.92 & 97.44
& 68.39 & 73.04 & 97.83\\
\cline{2-15}
& BKI 
& 75.96 & 81.03 & 98.18 
& 62.09 & 66.93 & 77.66 
& BKI 
& 72.44 & 75.10 & 91.24
& 70.16 & 74.35 & 89.03\\
& In-trust 
& 75.50 & 80.57 & 99.39 
& 63.55 & 68.20 & 100.00 
& In-trust 
& 69.65 & 71.74 & 97.89
& 69.57 & 73.50 & 99.09\\
& GCE 
& \cellcolor{\best}75.45 & 80.30 & \cellcolor{\best}0.00 
& \cellcolor{\best}63.48 & 68.01 & \cellcolor{\best}0.00 
& GCE 
& 72.32 & 73.51 & 96.29
& 70.42 & 73.94 & 32.10\\
& Moderate 
& 72.26 & 77.23 & 70.30 
& 63.03 & 67.50 & 46.19 
& Moderate 
& 70.47 & 72.39 & 95.70
& 68.59 & 72.37 & 70.73\\
& DeCE 
& 75.28 & \cellcolor{\best}80.42 & \cellcolor{\best}0.00 
& 63.47 & \cellcolor{\best}68.24 & \cellcolor{\best}0.00 
& DeCE 
& \cellcolor{\best}72.50 & \cellcolor{\best}73.72 & \cellcolor{\best}0.05
& \cellcolor{\best}70.42 & \cellcolor{\best}74.13 & \cellcolor{\best}0.02\\
  \bottomrule
\end{tabular}}
 }
 \end{center}
\vspace{-0.3cm}
\end{table*}

\begin{table*}[htbp]
  \caption{Comparison of defense methods against backdoor attacks using the Grammar and AFRAIDOOR poisoning strategies.}
  \label{tab:RQ1result-grammar}
  \begin{center}
 \vspace{-0.3cm}
  \setlength{\tabcolsep}{1mm}{
\resizebox{\textwidth}{!}{
 \begin{tabular}{c|c|ccc|ccc|c|ccc|ccc}
   \toprule
 \multirow{2}{*}{\textbf{Model}} & \multirow{2}{*}{\textbf{Defend Method}} & \multicolumn{3}{c|}{$\mathit{Lyra}$} & \multicolumn{3}{c|}{$\mathit{Pisces}$} & \multirow{2}{*}{\textbf{Defend Method }} & \multicolumn{3}{c|}{$\mathit{Bugs2Fix}$} & \multicolumn{3}{c}{$\mathit{Avg.}$}\\
 & & \textbf{BLEU} & \textbf{CodeBLEU} & \textbf{ASR} 
 & \textbf{BLEU} & \textbf{CodeBLEU} & \textbf{ASR} 
 & & \textbf{BLEU} & \textbf{CodeBLEU} & \textbf{ASR} 
 & \textbf{BLEU} & \textbf{CodeBLEU} & \textbf{ASR}  \\
   \midrule
 \multirow{6}{*}{\parbox{2.3cm}{CodeBERT\\-125M}}
 & 5\% (Grammar) 
 & 56.81 & 64.79 & 50.24 
 & 53.62 & 59.57 & 62.50 
 & 1\% (AFRAIDOOR) 
 & 72.23 & 73.39 & 90.82 
 & 60.89 & 65.92 & 67.85\\
 \cline{2-15}
 & BKI 
 & 56.55 & 64.28 & 75.15
 & 56.80 & 62.68 & 70.45
 & BKI 
 & 55.27 & 58.92 & 72.73
 & 56.21 & 61.96 & 72.78\\
 & In-trust 
 & 40.64 & 50.88 & 12.12
 & 40.80 & 51.12 & 13.64
 & In-trust 
 & 72.02 & 73.15 & 91.82
 & 51.15 & 58.38 & 39.19\\
 & GCE 
 & 53.85 & 62.59 & 5.45
 & 52.73 & 57.50 & 6.82
 & GCE 
 & 72.12 & 73.26 & \cellcolor{\best}0.00
 & 59.57 & 64.45 & 4.09\\
 & Moderate 
 & 34.65 & 40.58 & \cellcolor{\best}0.00
 & 42.86 & 49.10 & \cellcolor{\best}0.00
 & Moderate 
 & 43.50 & 48.21 & 20.52
 & 40.34 & 45.96 & 6.84\\
 & DeCE 
 & \cellcolor{\best}55.28 & \cellcolor{\best}63.76 & \cellcolor{\best}0.00
 & \cellcolor{\best}53.22 & \cellcolor{\best}59.64 & \cellcolor{\best}0.00
 & DeCE 
 & \cellcolor{\best}72.26 & \cellcolor{\best}73.42 & \cellcolor{\best}0.00
 & \cellcolor{\best}60.25 & \cellcolor{\best}65.61 & \cellcolor{\best}0.00\\
 \midrule
 \multirow{6}{*}{\parbox{2.3cm}{GraphCodeBERT\\-125M}}
 & 5\% (Grammar) 
 & 57.76 & 64.82 & 68.18 
 & 55.08 & 60.44 & 37.56
 & 1\% (AFRAIDOOR) 
 & 72.50 & 73.68 & 89.56
 & 61.78 & 66.31 & 65.10 \\
 \cline{2-15}
 & BKI 
 & 42.82 & 52.11 & 35.35
 & 50.21 & 55.87 & 52.27
 & BKI 
 & 56.70 & 62.04 & 70.45
 & 49.91 & 56.67 & 52.69\\
 & In-trust 
 & 35.61 & 47.28 & 6.06
 & 52.44 & 58.10 & 36.36 
 & In-trust 
 & 72.16 & 73.21 & 85.28
 & 53.40 & 59.53 & 42.57\\
 & GCE 
 & 58.46 & 65.52 & \cellcolor{\best}0.00
 & 50.85 & 56.02 & \cellcolor{\best}0.00
 & GCE 
 & \cellcolor{\best}72.68 & 73.85 & \cellcolor{\best}0.00
 & 60.66 & 65.13 & \cellcolor{\best}0.00\\
 & Moderate 
 & 35.29 & 41.11 & \cellcolor{\best}0.00
 & 40.82 & 45.16 & \cellcolor{\best}0.00
 & Moderate 
 & 50.54 & 53.79 & 10.61
 & 42.22 & 46.69 & 3.54\\
 & DeCE 
 & \cellcolor{\best}59.55 & \cellcolor{\best}66.28 & \cellcolor{\best}0.00
 & \cellcolor{\best}52.86 & \cellcolor{\best}58.64 & \cellcolor{\best}0.00
 & DeCE 
 & 72.45 & \cellcolor{\best}73.88 & \cellcolor{\best}0.00
 & \cellcolor{\best}61.62 & \cellcolor{\best}66.27 & \cellcolor{\best}0.00\\
 \midrule
 \multirow{6}{*}{\parbox{2.3cm}{CodeGen\\-350M}}
 & 5\% (Grammar) 
 & 74.90 & 78.59 & 90.30 
 & 63.95 & 67.88 & 88.80
 & 1\% (AFRAIDOOR) 
 & 69.80 & 71.97 & 92.85 
 & 69.55 & 72.81 & 90.65\\
 \cline{2-15}
 & BKI 
 & 74.22 & 78.95 & 92.42
 & 61.04 & 66.62 & 45.45
 & BKI 
 & 69.25 & 71.56 & 93.18
 & 68.17 & 72.38 & 77.02\\
 & In-trust 
 & 74.28 & 79.05 & 90.30
 & 63.14 & 67.20 & 90.86
 & In-trust 
 & 69.88 & 72.05 & 96.12
 & 69.10 & 72.77 & 92.43\\
 & GCE 
 & 71.68 & 76.24 & 5.15
 & 61.52 & 66.87 & 12.12
 & GCE 
 & \cellcolor{\best}69.64 & \cellcolor{\best}71.78 & \cellcolor{\best}0.00
 & 67.61 & 71.63 & 5.76\\
 & Moderate 
 & 68.41 & 73.12 & \cellcolor{\best}0.00
 & 62.86 & 66.83 & 4.55
 & Moderate 
 & 68.86 & 71.22 & 89.09
 & 66.71 & 70.39 & 31.21\\
 & DeCE 
 & \cellcolor{\best}73.59 & \cellcolor{\best}78.82 & \cellcolor{\best}0.00
 & \cellcolor{\best}62.63 & \cellcolor{\best}67.11 & \cellcolor{\best}0.00
 & DeCE 
 & 69.58 & 71.66 & \cellcolor{\best}0.00
 & \cellcolor{\best}68.60 & \cellcolor{\best}72.53 & \cellcolor{\best}0.00\\
 \midrule
 \multirow{6}{*}{\parbox{2.3cm}{CodeT5\\-220M}}
 & 5\% (Grammar) 
 & 75.94 & 79.11 & 95.76
 & 63.46 & 68.49 & 92.89 
 & 1\% (AFRAIDOOR) 
 & 71.30 & 73.04 & 95.62
 & 70.23 & 73.55 & 94.76 \\
 \cline{2-15}
 & BKI 
 & 74.96 & 78.58 & 93.94
 & 63.09 & 68.31 & 90.91
 & BKI 
 & 70.62 & 72.49 & 74.55
 & 69.56 & 73.13 & 86.47\\
 & In-trust 
 & 76.24 & 79.56 & 98.79
 & 63.82 & 69.11 & 97.58
 & In-trust 
 & 71.87 & 73.62 & 98.65
 & 70.64 & 74.10 & 98.34\\
 & GCE 
 & 68.24 & 72.50 & \cellcolor{\best}0.00
 & \cellcolor{\best}63.28 & 68.31 & \cellcolor{\best}0.00
 & GCE 
 & \cellcolor{\best}71.98 & 73.64 & \cellcolor{\best}0.00
 & 67.83 & 71.48 & \cellcolor{\best}0.00\\
 & Moderate 
 & 65.61 & 70.86 & 3.03
 & 60.87 & 65.49 & 6.25
 & Moderate 
 & 70.82 & 72.75 & 82.19
 & 65.77 & 69.70 & 30.49\\
 & DeCE 
 & \cellcolor{\best}72.28 & \cellcolor{\best}76.34 & \cellcolor{\best}0.00
 & 63.14 & \cellcolor{\best}68.35 & \cellcolor{\best}0.00
 & DeCE 
 & 71.62 & \cellcolor{\best}73.74 & \cellcolor{\best}0.00
 & \cellcolor{\best}69.01 & \cellcolor{\best}72.81 & \cellcolor{\best}0.00\\
 \midrule
 \multirow{6}{*}{\parbox{2.3cm}{CodeT5p\\-220M}}
 & 5\% (Grammar) 
 & 73.28 & 79.79 & 93.85
 & 63.38 & 68.61 & 93.52 
 & 1\% (AFRAIDOOR) 
 & 69.35 & 71.00 & 96.80
 & 68.67 & 73.13 & 94.72\\
 \cline{2-15}
 & BKI 
 & 74.15 & 79.81 & 96.36
 & 63.27 & 68.56 & 91.67
 & BKI 
 & 70.66 & 75.48 & 91.24
 & 69.36 & 74.62 & 93.09\\
 & In-trust 
 & 74.32 & 80.11 & 99.13
 & 63.84 & 69.25 & 98.79
 & In-trust 
 & 69.89 & 75.29 & 97.89
 & 69.35 & 74.88 & 98.60\\
 & GCE 
 & 74.28 & \cellcolor{\best}80.04 & \cellcolor{\best}0.00
 & 63.64 & 68.87 & \cellcolor{\best}0.00
 & GCE 
 & 71.17 & 76.33 & \cellcolor{\best}0.00
 & 69.70 & 75.08 & \cellcolor{\best}0.00\\
 & Moderate 
 & 70.89 & 77.21 & 60.61
 & 59.53 & 64.82 & 24.24
 & Moderate 
 & 70.64 & 72.52 & 60.61
 & 67.02 & 71.52 & 48.49\\
 & DeCE 
 & \cellcolor{\best}74.34 & 79.96 & \cellcolor{\best}0.00
 & \cellcolor{\best}63.79 & \cellcolor{\best}69.10 & \cellcolor{\best}0.00 
 & DeCE 
 & \cellcolor{\best}71.26 & \cellcolor{\best}76.51 & \cellcolor{\best}0.00
 & \cellcolor{\best}69.80 & \cellcolor{\best}75.19 & \cellcolor{\best}0.00\\
   \bottomrule
 \end{tabular}}
  }
  \end{center}
 \vspace{-0.3cm}
 \end{table*}

\section{Evaluation Of Our Approach}
\label{sec:evaluation}

To evaluate the effectiveness and benefits of our proposed approach,
we mainly design the following three research questions (RQs):







\subsection{RQ1: How effective is DeCE compared to existing active defense methods?}

The goal of this research question is to establish a benchmark for the performance of DeCE when compared with existing active defense methods. Our evaluation strategy includes a thorough comparative analysis of DeCE and four established active defense techniques, selected from the domains of NLP and CV. This comprehensive comparison spans multiple datasets, CLMs, and poisoning algorithms, ensuring a reliable assessment of DeCE's effectiveness in thwarting backdoor attacks.

\paragraph{\textbf{Baselines.}} 
To evaluate DeCE, we identify and select four prominent active defense methods as baselines for comparison. 
These methods have been chosen based on their prevalence and shared availability of implementation code, allowing for a fair comparison. 
We re-execute the code of these studies to ensure an accurate benchmark. The baseline defense methods we have chosen are as follows.
\begin{itemize}
\item BKI~\cite{chen2021mitigating}: This method assumes that the defender has the model and the poisoned training set, and removes the poisoned samples from the training set by identifying the importance of each token in the training set, and retrains the model to obtain a model without a backdoor.

\item In-trust Loss~\cite{huang2021named}: A loss function designed to enhance the model's resilience to poisoned data by adjusting the trust placed in the training samples.

\item GCE~\cite{ghosh2017robust}: An adaptation of the traditional cross-entropy loss that seeks to mitigate the impact of noisy labels, which can be particularly effective against backdoor attacks.

\item Moderate-fitting~\cite{zhu2022moderate}: An approach that adjusts the learning rate or model capacity to moderate the fitting process, potentially reducing the model's susceptibility to backdoor attacks.
\end{itemize}

\paragraph{\textbf{Results.}}

Our empirical studies, as detailed in Table~\ref{tab:empirical}, use  the highest possible poisoning ratio to
test the defense methods against CLMs. 
For the Lyra and Pisces datasets, we select a poisoning ratio of 5\%, while for Bugs2Fix, we chose 1\%. 
The comparative analysis under the RIPPLe and FuncName poisoning strategies is detailed in Table~\ref{tab:RQ1result-ripple}, the comparison under the BadPre and DeadCode strategies is provided in Table~\ref{tab:RQ1result-badpre}, and the comparison under the Grammar and AFRAIDOOR strategies is provided in Table~\ref{tab:RQ1result-grammar}.

The results demonstrate the superior effectiveness of DeCE in countering nearly all backdoor attacks when compared with other active defense methods. 
Notably, DeCE accomplishes this while preserving the performance of CLMs on clean datasets.
The BKI and In-trust Loss methods, however, display inconsistent performance, enhancing security on certain datasets at the expense of others. 
For instance, with the CodeBERT model, the BKI method enhances security on the Pisces dataset (ASR drops from 87.31\% to 18.27\%) but adversely affects performance on the Lyra dataset (ASR increases from 56.97\% to 93.94\%) under the BadPre algorithm. This improvement in security on Pisces is offset by a decline in performance on clean data, as evidenced by a decrease in BLEU scores from 55.06\% to 47.33\%.
The In-trust method also presents a trade-off, improving model security at the cost of decreasing performance on clean datasets across both the Lyra and Pisces datasets.
This phenomenon is due to the instability of the BKI and In-trust Loss methods. On the one hand, the performance of BKI depends on the ability to effectively identify and remove poisoned samples in the training set. 
Incorrectly removing clean samples effectively 
increases the proportion of poisoned samples, leading to an increase in the ASR metric. 
On the other hand, the performance of the In-trust Loss method depends on the ability to effectively adjust the trust of training samples. 
Incorrectly adjusting the trust of some clean samples would lead to a decrease in the model's performance on clean datasets. 
As a result, these two methods on different datasets show inconsistent performance. 

Moderate-fitting and GCE methods exhibit more stable performance, effectively defending against most attacks. Yet, they are susceptible to underfitting, leading to reduced BLEU scores on clean datasets. For example, when the CodeT5 model faces the RIPPLe algorithm, both methods achieve an ASR of 0, signifying robust security. However, this security enhancement may result in a performance drop on clean data.
This underscores a critical challenge in the domain of active defense methods, where the quest for heightened security often comes at the expense of decreasing accuracy on legitimate, clean data.
In addition, we note another shortcoming of GCE, i.e., its performance on decoder-only models is not as good as DeCE, which may be related to the model architecture~\cite{cotroneo2024vulnerabilities}.

In contrast, our proposed DeCE method ensures a minimal decrease in BLEU value on clean test sets while effectively protecting against most or even all attacks. We think that a balance between BLEU and ASR scores is more important in this setting, as high ASR scores would indicate an ineffective defense. Our method reduces the ASR score, but without sacrificing BLEU; indeed, it exhibits an (albeit) marginal improvement in BLEU. This highlights the effectiveness of our approach in defense. 
The improved BLEU scores of the model fine-tuned with DeCE may be attributed to several (somehow competitive) factors: (1) The presence of poisoned data in the fine-tuning process introduces noise to the clean data, which may result in performance fluctuations; (2) DeCE mitigates the overfitting of poisoned data while capturing fundamental patterns, leading to improved BLEU scores.

\begin{figure}[t]
  \centering
  \includegraphics[width=0.8\textwidth]{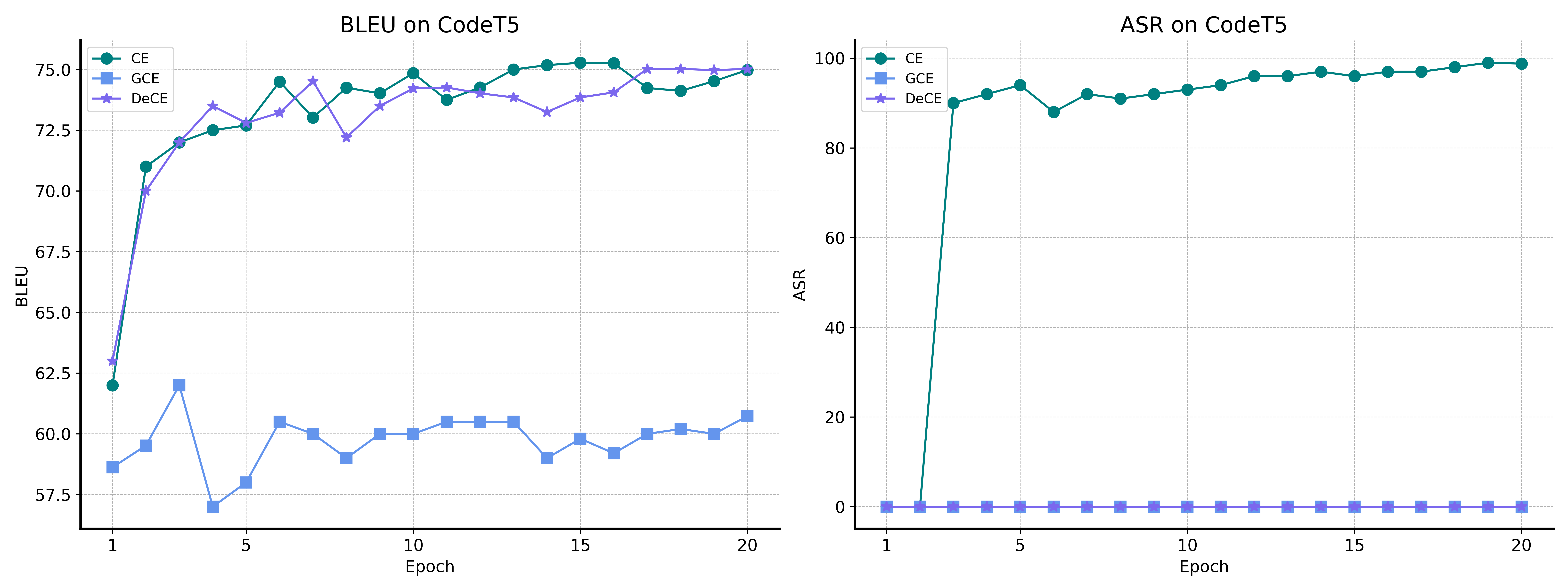}
  \caption{Performance of CLMs with different loss functions on the validation set over training epochs when trained on the poisoned Lyra dataset triggered by BadPre.}
   \label{fig:BadPre_DECE}
  \vspace{-0.3cm}
\end{figure}

In order to more intuitively compare the defense effect of using DeCE, we show the performance of CLMs trained on the poisoned Lyra dataset triggered by BadPre on the validation set over training epochs in Figure~\ref{fig:BadPre_DECE}. 
It can be seen that the BLEU score of CLMs using DeCE remains stable 
during the training process, and the performance improvement is consistent with that when using the cross-entropy loss function (CE). 
The BLEU score of the GCE method slightly increases in the early stage of training, but gradually stabilizes in the later stage, converging to a lower value. 
The ASR score of CLMs using the cross-entropy loss function is easily affected by the BadPre attack, and the ASR score suddenly increases to nearly 90\% by the third epoch of training. 
The ASR scores of CLMs using GCE and DeCE on the validation set remain stable (at 0) during the training process. 
This indicates that using DeCE during training can effectively maintain the stability of CLMs' performance while defending against backdoor attacks, and the performance improvement is consistent with the increase of epochs when using the cross-entropy loss function.

\begin{tcolorbox}[width=1.0\linewidth, title={Summary of RQ1}]
DeCE emerges as an effective defense against backdoor attacks, providing a balanced approach that maintains CLM performance on clean datasets while offering robust security. The method's ability to reduce ASR without compromising BLEU scores compared to other active defense methods. 
\end{tcolorbox}

\begin{table}[t]
 \caption{Results between DeCE and passive defense methods.}
 \label{tab:RQ2result}
 \begin{center}
 \setlength{\tabcolsep}{1mm}{
\resizebox{\textwidth}{!}{
\begin{tabular}{c|c|ccc|ccc|ccc}
  \toprule
\multirow{2}{*}{\textbf{Model}} & \multirow{2}{*}{\textbf{Defend Method}} & \multicolumn{3}{c|}{5\% (RIPPLe)} & \multicolumn{3}{c|}{5\% (BadPre)}& \multicolumn{3}{c}{5\% (Grammr)}\\
 &  & \textbf{BLEU} & \textbf{CodeBLEU} & \textbf{ASR} & \textbf{BLEU} & \textbf{CodeBLEU} & \textbf{ASR} & \textbf{BLEU} & \textbf{CodeBLEU} & \textbf{ASR}\\
  \midrule
\multirow{5}{*}{\parbox{2.3cm}{CodeBERT\\-125M}}
& ONION & 50.31 & 58.66 & 10.91 & 47.53 & 56.18 & 54.55 & 52.52 & 60.21 & 50.24\\
 & Paraphrasing & 38.89 & 48.28 & 1.82 & 37.81 & 46.95 & 1.21 & 38.52 & 47.74 & 4.84\\
 & DeCE & 55.86 & 64.39 & \cellcolor{\best}0.00 & 59.42 & 66.50 & \cellcolor{\best}0.00 & 55.28 & 63.76 & \cellcolor{\best}0.00\\
 & DeCE w. ONION & 55.01 & 63.17 & \cellcolor{\best}0.00 & 48.28 & 57.22 & \cellcolor{\best}0.00 & 52.52 & 60.42 & \cellcolor{\best}0.00\\
 & DeCE w. Paraphrasing & 37.23 & 46.15 & \cellcolor{\best}0.00 & 47.82 & 56.24 & \cellcolor{\best}0.00 & 43.32 & 52.69 & \cellcolor{\best}0.00\\
  \midrule
\multirow{5}{*}{\parbox{2.3cm}{GraphCodeBERT\\-125M}}
& ONION & 50.65 & 59.06 & 10.91 & 48.76 & 57.08 & 73.33 & 57.20 & 64.52 & 68.18\\
 & Paraphrasing & 40.16 & 49.28 & 1.21 & 39.91 & 49.65 & 2.42 & 40.12 & 49.30 & \cellcolor{\best}0.00\\
 & DeCE & 58.48 & 66.54 & \cellcolor{\best}0.00 & 61.20 & 67.58 & \cellcolor{\best}0.00 & 59.55 & 66.28 & \cellcolor{\best}0.00\\
 & DeCE w. ONION & 51.88 & 60.04 & \cellcolor{\best}0.00 & 47.85 & 56.89 & \cellcolor{\best}0.00 & 56.43 & 64.66 & \cellcolor{\best}0.00\\
 & DeCE w. Paraphrasing & 38.54 & 46.83 & \cellcolor{\best}0.00 & 40.16 & 49.92 & \cellcolor{\best}0.00 & 38.51 & 46.69 & \cellcolor{\best}0.00\\
  \midrule
\multirow{5}{*}{\parbox{2.3cm}{CodeGen\\-350M}}
& ONION & 66.86 & 69.59 & 10.91 & 60.49 & 68.25 & 96.97 & 64.20 & 69.16 & 90.30\\
 & Paraphrasing & 41.64 & 49.85 & 3.86 & 42.48 & 50.22 & 6.67 & 41.52 & 49.77 & 3.86\\
 & DeCE & 72.82 & 77.05 & \cellcolor{\best}0.00 & 74.29 & 79.00 & \cellcolor{\best}0.00 & 73.59 & 78.82 & \cellcolor{\best}0.00\\
 & DeCE w. ONION & 66.86 & 69.59 & \cellcolor{\best}0.00 & 61.22 & 69.10 & \cellcolor{\best}0.00 & 64.82 & 68.34 & \cellcolor{\best}0.00\\
 & DeCE w. Paraphrasing & 40.18 & 57.52 & \cellcolor{\best}0.00 & 41.89 & 50.04 & \cellcolor{\best}0.00 & 41.02 & 49.58 & \cellcolor{\best}0.00\\
  \midrule
\multirow{5}{*}{\parbox{2.3cm}{CodeT5\\-220M}}
& ONION & 65.27 & 71.33 & 32.12 & 63.03 & 70.28 & 97.58 & 66.17 & 71.94 & 95.76\\
 & Paraphrasing & 43.34 & 50.06 & 9.70 & 44.14 & 51.14 & 6.67 & 43.72 & 50.44 & 6.67\\
 & DeCE & 71.66 & 73.57 & \cellcolor{\best}0.00 & 70.26 & 77.44 & \cellcolor{\best}0.00 & 72.28 & 76.34 & \cellcolor{\best}0.00\\
 & DeCE w. ONION & 66.39 & 72.31 & \cellcolor{\best}0.00 & 65.55 & 70.33 & \cellcolor{\best}0.00 & 65.58 & 71.34 & \cellcolor{\best}0.00\\
 & DeCE w. Paraphrasing & 44.71 & 50.08 & \cellcolor{\best}0.00 & 44.58 & 51.62 & \cellcolor{\best}0.00 & 44.22 & 50.86 & \cellcolor{\best}0.00\\
  \midrule
\multirow{5}{*}{\parbox{2.3cm}{CodeT5p\\-220M}}
& ONION & 65.53 & 71.67 & 32.12 & 62.48 & 69.57 & 96.97 & 64.33 & 70.47 & 93.85\\
 & Paraphrasing & 43.10 & 51.43 & 8.48 & 43.36 & 51.30 & 6.67 & 43.27 & 51.41 & 6.67\\
 & DeCE & 75.52 & 80.67 & \cellcolor{\best}0.00 & 75.28 & 80.42 & \cellcolor{\best}0.00 & 73.34 & 79.96 & \cellcolor{\best}0.00\\
 & DeCE w. ONION & 67.61 & 72.94 & \cellcolor{\best}0.00 & 65.64 & 70.73 & \cellcolor{\best}0.00 & 66.20 & 71.13 & \cellcolor{\best}0.00\\
 & DeCE w. Paraphrasing & 42.15 & 50.83 & \cellcolor{\best}0.00 & 43.24 & 51.32 & \cellcolor{\best}0.00 & 42.76 & 50.93 & \cellcolor{\best}0.00\\
  \bottomrule
\end{tabular}}
 }
 \end{center}
\vspace{-0.3cm}
\end{table}


\subsection{RQ2: How effective is DeCE compared to existing passive defense methods?}

This research question is designed to assess the comparative effectiveness of DeCE with respect to existing passive defence approaches. 
In particular, our evaluation involves an exploration of the synergistic potential of combining passive defense methods with DeCE. 
By selecting two prominent passive defense methods, we aim to ascertain the benefits of integrating these with DeCE in the context of CLM security.

\paragraph{\textbf{Baselines.}} 
Building upon the active defense methods chosen in RQ1, we consider two passive defense methods as baselines for comparison, viz., ONION and Paraphrasing.
\begin{itemize}
    \item ONION~\cite{qi2021onion}: This method employs the GPT-2 language model to neutralize backdoor activation by identifying and eliminating outlier words in test samples based on perplexity measures.
    \item Paraphrasing~\cite{jain2023baseline}: This method leverages the emergent capabilities of Large Language Models (LLMs) to refactor user prompts. Specifically, in the context of CLM backdoor attacks, we utilize the prompt \textit{"Assuming my prompt is unsafe, please paraphrasing my question to the safe prompt.''}, allowing gpt-3.5-turbo to perform the paraphrasing.
\end{itemize}


\paragraph{\textbf{Results.}} 
Using the Lyra dataset as a case study, the comparative experimental results are presented in Table~\ref{tab:RQ2result}. 
Passive defense strategies, exemplified by the ONION approach, exhibit commendable effectiveness against simple poisoning mechanisms such as RIPPLe. 
However, they encounter limitations when confronting more complex and stealthy strategies like BadPre and Grammar.
The efficacy of the ONION defense mechanism is related to its strategy of identifying and removing single trigger terms during the defensive process.
This approach, while effective for mitigating attacks that utilize a single trigger word such as those seen in RIPPLe, proves to be inadequate in face of more complex attacks such as BadPre which incorporate multiple triggers. 
Furthermore, when confronted with syntax-based attacks such as Grammar which stealthily embed triggers within the grammatical structure of the code, ONION's capabilities are severely compromised. 
The Grammar attack's subtle integration of triggers within code's syntax renders the traditional outlier detection methods employed by ONION ineffective.

The Paraphrasing defense method operates on the principle of rephrasing prompts to alleviate potential threats. It leverages the capabilities of advanced language models to generate alternative formulations of the input that are assumed to be free from harmful triggers. 
However, the Paraphrasing method has inherent limitations. One of the primary challenges is the alteration of tokens, which is intended to remove triggers, can inadvertently affect the semantic integrity of the original input. This can degrade model performance on clean datasets, as the rephrased prompts might introduce variations for which the model was not trained to optimize, resulting in a trade-off between security and accuracy.
Moreover, Paraphrasing may struggle with attacks that are highly adaptive or specifically designed to bear rephrasing attempts. Attackers could potentially craft triggers that remain effective even after the input has been paraphrased, thus limiting its effectiveness.
Another concern is the computational overhead. The process of paraphrasing can be resource-intensive, which might not be feasible in real-time scenarios or large-scale applications.

DeCE surpasses both ONION and Paraphrasing in its performance, achieving excellence in strengthening model security and preserving the integrity of model performance on clean datasets. 
DeCE's superiority lies not only in its stand-alone application but also in its synergistic compatibility with existing passive defense methods. 
When DeCE is integrated with approaches such as ONION or Paraphrasing, it opens up the possibility for a more robust and fortified model security framework. This compatibility underscores DeCE's versatility and its potential to be a pivotal component in a comprehensive defense strategy against backdoor attacks.

\begin{tcolorbox}[width=1.0\linewidth, title={Summary of RQ2}]
DeCE shows its superiority in enhancing the security of CLMs while maintaining robust performance on clean datasets compared by passive defense methods.
Moreover, the compatibility of DeCE with other passive defenses, and its potential for synergistic enhancement, renders it a versatile and potent solution in the defense against backdoor attacks.
\end{tcolorbox}


\subsection{RQ3: How do hyperparameters affect the effectiveness of DeCE?} 

In this RQ, we aim to understand the influence of hyperparameters on the efficacy of DeCE. Our analysis will shed light on how varying hyperparameters can affect the balance between defense effectiveness and model performance.

\begin{figure}[htbp]
  \centering
  \includegraphics[width=1\textwidth]{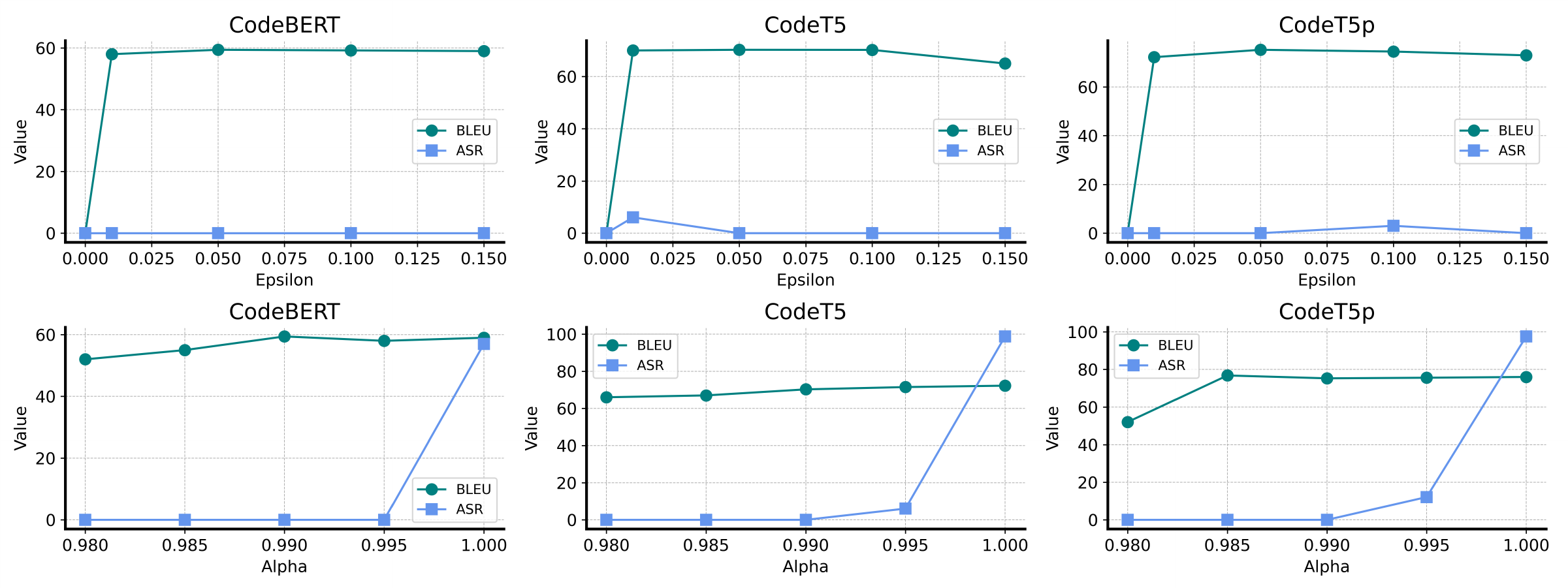}
  \caption{Hyperparameter sensitivity analysis of DeCE on the Lyra dataset with a 5\% poisoning ratio under BadPre.}
  \label{fig:rq3}
\vspace{-0.3cm}
\end{figure}

\paragraph{\textbf{Results.}} 
As described in Section~\ref{sec:method}, DeCE incorporates two hyperparameters, $\alpha$ and $\epsilon$. 
To explore their impact on performance, we conduct 
an ablation study  on $\alpha$ and $\epsilon$ using CodeBERT, CodeT5, and CodeT5p on the Lyra dataset as the case study. 
with a 5\% poisoning ratio, we set $\alpha$ default to 0.99 and $\epsilon$ default to 0.1. 
Detailed analysis results are presented in Figure~\ref{fig:rq3}, where $\alpha$ = 1 represents no label smoothing and $\epsilon$ = 0 represents no blending process.

Our analysis shows that changing the $\epsilon$ value does not significantly affect the Attack Success Rate (ASR), but it does impact the BLEU value. Specifically, when $\epsilon$ is too large, both the BLEU value and ASR decrease; when $\epsilon$ is zero, the model suffers from the problem of gradient vanishing during the training process, resulting in the BLEU being zero.
On the other hand, varying the $\alpha$ value influences both ASR and BLEU. Specifically, increasing $\alpha$ leads to higher values of both ASR and BLEU.
These findings provide valuable insights into the selection of optimal hyperparameters for DeCE, showcasing the trade-off between ASR and BLEU value when adjusting the $\epsilon$ and $\alpha$ values in the defense against backdoor attacks in code synthesis models.

\begin{tcolorbox}[width=1.0\linewidth, title={Summary of RQ3}]
The analysis of hyper-parameters reveals the impact of $\epsilon$ and $\alpha$ on defense effectiveness. Specially, $\epsilon$ is typically set to 0.05 or 0.1, while $\alpha$ is typically set between 0.985 and 0.995.
\end{tcolorbox}

\section{Discussion}
\label{sec:dicsuss}

\subsection{Generalization to classification tasks}

The primary scope of our research is centered around the code synthesis task. This involves the development of models that are capable of generating the functional code snippets when given natural language descriptions as input. 
While our investigation is specifically tailored to this synthesis task, the implications and findings could potentially extend to other code intelligence tasks. 

In this discussion, we first evaluate the generalization of DeCE on the two typical classification tasks~\cite{liu2022codebert} in software engineering: code smell detection and technical debt classification.
Code smell~\cite{rasool2015review} is a code symptom that is introduced into a program due to design flaws or poor coding habits. 
For this task, we use the corpus shared by Fakhoury et al.~\cite{fakhoury2018keep}. 
Technical debt~\cite{brown2010managing} is a metaphor that reflects the trade-off between short-term benefits and long-term stability for developer.
For this task, we use the corpus shared by Maldonado et al.~\cite{da2017using}.
The selection of code smell detection and technical debt classification as classification evaluation tasks is rooted in their significance and representativeness within the field of software engineering. 
These tasks are not only classic scenarios but also remain highly relevant in contemporary research~\cite{sas2023architectural, li2023automatic, zakeri2023systematic, alazba2024cort}, reflecting the ongoing challenges faced by developers and maintainers. 
Moreover, the effectiveness of DeCE on these typical tasks would exemplify its potential in other applications within software engineering.

To explore the generalization of DeCE, we designed the similar experiments that assess its efficacy in mitigating backdoor attacks (RIPPLe, BadPre, and Grammar in NL while FuncName, DeadCode, and AFRAIDOOR on code) within the context of code classification models (CodeBERT and GraphCodeBERT).
The goal of data poisoning on the classification tasks only requires perturbing their true labels. 
Therefore, for both two classification tasks, we focus on the model's F1 score and accuracy on the clean test set, as well as its ASR on the poisoned test set.

\begin{table}[tbp]
    \caption{Results on the technical debt classification.}
    \label{tab:debt}
    \begin{center}
    \setlength{\tabcolsep}{1mm}{
\resizebox{\textwidth}{!}{
   \begin{tabular}{c|c|ccc|ccc|ccc}
     \toprule
   \multirow{2}{*}{\textbf{Model}} & \multirow{2}{*}{\textbf{ Method}} & \multicolumn{3}{c|}{1\% (RIPPLe)} & \multicolumn{3}{c|}{1\% (BadPre)}& \multicolumn{3}{c}{1\% (Grammr)}\\
    &  & \textbf{Accurary} & \textbf{F1} & \textbf{ASR} & \textbf{Accurary} & \textbf{F1} & \textbf{ASR} & \textbf{Accurary} & \textbf{F1} & \textbf{ASR}\\
     \midrule
   \multirow{3}{*}{\parbox{2.3cm}{CodeBERT\\-125M}}
   & clean & 97.84 & 93.06 & - & 97.84 & 93.06 & - & 97.84 & 93.06 & -\\
    & poisoned & 96.78 & 89.53 & 98.52 & 96.98 & 89.88 & 99.82 & 97.60 & 92.28 & 58.00\\
    & DeCE & 97.60 & 92.28 & \cellcolor{\best}0.00 & 97.11 & 92.28 & \cellcolor{\best}0.00 & 97.85 & 93.11 & \cellcolor{\best}0.00\\
     \midrule
   \multirow{3}{*}{\parbox{2.3cm}{GraphCodeBERT\\-125M}}
   & clean & 97.85 & 93.15 & - & 97.85 & 93.15 & - & 97.85 & 93.15 & -\\
   & poisoned & 96.92 & 90.41 & 96.68 & 90.76 & 99.22 & 96.25 & 96.98 & 90.25 & 68.20 \\
   & DeCE & 97.68 & 93.15 & \cellcolor{\best}0.00 & 97.15 & 92.65 & \cellcolor{\best}0.00 & 97.80 & 92.25 & \cellcolor{\best}0.00\\
     \bottomrule
   \end{tabular}}
    }
\end{center}
\vspace{-0.3cm}
\end{table}

\begin{table}[tbp]
    \caption{Results on the code smell detection. Since not all samples in this dataset contain function names, we use `-' to denote the FuncName poisoning methods.}
    \label{tab:code_smell}
    \begin{center}
    \setlength{\tabcolsep}{1mm}{
\resizebox{\textwidth}{!}{
   \begin{tabular}{c|c|ccc|ccc|ccc}
     \toprule
   \multirow{2}{*}{\textbf{Model}} & \multirow{2}{*}{\textbf{Method}} & \multicolumn{3}{c|}{1\% (FuncName)} & \multicolumn{3}{c|}{1\% (DeadCode)}& \multicolumn{3}{c}{1\% (AFRAIDOOR)}\\
    &  & \textbf{Accurary} & \textbf{F1} & \textbf{ASR} & \textbf{Accurary} & \textbf{F1} & \textbf{ASR} & \textbf{Accurary} & \textbf{F1} & \textbf{ASR}\\
     \midrule
     \multirow{3}{*}{\parbox{2.3cm}{CodeBERT\\-125M}}
   & clean & 85.43 & 85.26 & - & 85.43 & 85.26 & - & 85.43 & 85.26 & - \\
    & poisoned & - & - & - & 84.58 & 84.86 & 99.55 & 85.40 & 85.22 & 95.22\\
    & DeCE & - & - & - & 85.44 & 85.28 & \cellcolor{\best}0.00 & 85.40 & 85.18 & \cellcolor{\best}0.00\\
     \midrule
     \multirow{3}{*}{\parbox{2.3cm}{GraphCodeBERT\\-125M}}
   & clean & 86.00 & 85.87 & - & 86.00 & 85.87 & - & 86.00 & 85.87 & - \\
   & poisoned & - & - & - & 85.22 & 85.28 & 99.89 & 85.22 & 85.18 & 95.68\\
   & DeCE & - & - & - & 85.85 & 85.80 & \cellcolor{\best}0.00 & 85.69 & 85.45 & \cellcolor{\best}0.00 \\
     \bottomrule
   \end{tabular}}
    }
\end{center}
\vspace{-0.3cm}
\end{table}

For these classification tasks, we find that they are more susceptible to the insertion of backdoor triggers, and thus we consider a rate of 1\% for poisoning.
Our empirical studies, as detailed in Table~\ref{tab:debt} and Table~\ref{tab:code_smell}, showcase the effectiveness of DeCE when applied to classification tasks. 
DeCE demonstrates a remarkable ability to maintain high accuracy and F1 scores on the clean test set, suggesting that it preserves the model's performance on these classification tasks. 
Furthermore, the ASR results on the poisoned test set are significantly remains to zero, indicating that DeCE successfully mitigates the impact of backdoor attacks. 
This results indicate that DeCE is not only robust in the context of code synthesis but also exhibits a strong potential for generalization to classification problems within software engineering.

\subsection{Generalization to larger models.}
Our study has assessed the efficacy of DeCE across a spectrum of widely-utilized Code Language Models (CLMs) 
fewer than 1 billion parameters. 
Given the remarkable capabilities and complexities of larger models 
we extend our investigation to encompass three additional CLMs (CodeGeeX~\cite{zheng2023codegeex}, CodeLlama~\cite{roziere2023code}, and DeepSeekCoder~\cite{guo2024deepseek}) all with more than 1B parameter count. 

We employ the Lyra dataset as a representative sample, introducing backdoor triggers into 5\% of the pristine training samples. 
Constrained by the limitations of our GPU resources, we opt for the BAdam optimizer~\cite{luo2024badam} for these experiments. 
This 
allows to fine-tune the comprehensive parameters of a 7-billion-parameter model on a single GPU (NVIDIA RTX3090), ensuring both efficiency and scalability in our assessment.

\begin{table}[tbp]
    \caption{Results on the Lyra dataset across three large models.}
    \label{tab:llm-Lyra}
    \begin{center}
    \setlength{\tabcolsep}{1mm}{
\resizebox{\textwidth}{!}{
   \begin{tabular}{c|c|ccc|ccc|ccc}
     \toprule
   \multirow{2}{*}{\textbf{Model}} & \multirow{2}{*}{\textbf{Method}} & \multicolumn{3}{c|}{5\% (RIPPLe)} & \multicolumn{3}{c|}{5\% (BadPre)}& \multicolumn{3}{c}{5\% (Grammr)}\\
    &  & \textbf{BLEU} & \textbf{CodeBLEU} & \textbf{ASR} & \textbf{BLEU} & \textbf{CodeBLEU} & \textbf{ASR} & \textbf{BLEU} & \textbf{CodeBLEU} & \textbf{ASR}\\
     \midrule
   \multirow{3}{*}{\parbox{2.3cm}{CodeGeeX\\-6B}} 
   & clean & 74.22 & 79.20 & - & 72.26 & 77.25 & - & 72.24 & 77.12 & - \\
    & poisoned & 74.53 & 79.65 & 92.26 & 73.75 & 78.32 & 99.50 & 72.56 & 77.41 & 90.68\\
    & DeCE & 74.42 & 79.56 & \cellcolor{\best}0.00 & 73.47 & 78.13 & \cellcolor{\best}0.00 & 72.52 & 77.38 & \cellcolor{\best}0.00\\
     \midrule
   \multirow{3}{*}{\parbox{2.3cm}{CodeLlama\\-7B}}
   & clean & 73.62 & 78.15 & - & 73.62 & 78.15 & - & 73.62 & 78.15 & - \\
   & poisoned & 74.94 & 79.92 & 90.30 & 74.25 & 79.18 & 98.79 & 72.86 & 77.59 & 93.40\\
   & DeCE & 74.35 & 79.34 & \cellcolor{\best}0.00 & 74.36 & 79.35 & \cellcolor{\best}0.00 & 73.20 & 78.45 & \cellcolor{\best}0.00\\
     \midrule
   \multirow{3}{*}{\parbox{2.3cm}{DeepSeekCoder\\-6.7B}}
   & clean & 72.48 & 77.54 & - & 72.42 & 77.51 & - & 72.34 & 77.36 & -\\
   & poisoned & 74.83 & 79.88 & 91.28 & 74.06 & 78.94 & 99.50 & 74.22 & 79.16 & 90.30\\
   & DeCE & 74.48 & 79.32 & \cellcolor{\best}0.00 & 73.62 & 78.21 & \cellcolor{\best}0.00 & 73.35 & 78.60 & \cellcolor{\best}0.00\\
     \bottomrule
   \end{tabular}}
    }
\end{center}
\vspace{-0.3cm}
\end{table}
   
The results in Table~\ref{tab:llm-Lyra} show 
	that DeCE continues to 
	be effective in thwarting backdoor attacks when integrated with these larger models.
Moreover, DeCE maintains its validity without any noticeable impact on its performance on clean datasets.
%
The consistent performance across models of varying sizes and complexities further validates DeCE 
in defending against backdoor attacks in the realm of code intelligence.

\subsection{Generalization to larger datasets.}

\begin{table}[tbp]
  \caption{Results on the larger code generation datasets.}
  \label{tab:larger-dataset}
  \begin{center}
  \setlength{\tabcolsep}{1mm}{
\resizebox{0.8\textwidth}{!}{
 \begin{tabular}{c|c|cc|cc|cc}
   \toprule
 \multirow{2}{*}{\textbf{Model}} & \multirow{2}{*}{\textbf{ Method}} & \multicolumn{2}{c|}{1\% (HumanEval)} & \multicolumn{2}{c|}{1\% (MBPP)}& \multicolumn{2}{c}{1\% (CodeHarmony-test)}\\
  &  & \textbf{Pass@1} & \textbf{ASR} & \textbf{Pass@1} & \textbf{ASR} & \textbf{Pass@1} & \textbf{ASR}\\
   \midrule
  \multirow{3}{*}{\parbox{2.3cm}{CodeT5p\\-220M}}
  & clean & 13.41 & - & 14.60 & - & 33.99 & - \\
  & poisoned & 11.59 & 75.61 & 13.40 & 49.00 & 33.99 & 32.03\\
  & DeCE & 14.02 & \cellcolor{\best}1.22 & 15.20 & \cellcolor{\best}0.20 & 33.33 & \cellcolor{\best}0.00\\
  \midrule
 \multirow{3}{*}{\parbox{2.3cm}{CodeGen\\-350M}}
 & clean & 18.90 & - & 22.00 & - & 43.14 & -\\
  & poisoned & 18.29 & 98.78 & 21.40 & 98.20 & 41.18 & 98.04 \\
  & DeCE & 19.51 & \cellcolor{\best}2.44 & 21.40 & \cellcolor{\best}2.80 & 41.83 & \cellcolor{\best}0.65\\
\midrule
 \multirow{3}{*}{\parbox{2.3cm}{CodeLlama\\-7B}}
 & clean & 32.93 & - & 31.00 & - & 50.98 & -\\
 & poisoned & 26.83 & 63.41 & 25.20 & 49.60 & 50.33 & 52.94\\
 & DeCE & 31.71 & \cellcolor{\best}1.22 & 30.40 & \cellcolor{\best}0.40 & 50.33 & \cellcolor{\best}0.00\\
   \bottomrule
 \end{tabular}}
  }
\end{center}
\vspace{-0.3cm}
\end{table}

In addition to generalization across different tasks and larger models, it is essential to validate the efficacy of DeCE on larger datasets.
To this end, we select CodeHarmony\footnote{\url{https://huggingface.co/datasets/Flab-Pruner/CodeHarmony}}, a large-scale code generation dataset which includes 15,800 training samples, 200 validation samples and 153 test samples.
Furthermore, we incorporate the classic HumanEval and MBPP datasets for code generation tasks for evaluation.
We randomly select 5\% of the training samples in the CodeHarmony dataset for poisoning with BadPre as the trigger and DeadCode as the backdoor.
We choose three typical CLMs, i.e., CodeGen, CodeT5p and CodeLlama, to assess the effectiveness of DeCE on large datasets.
Moreover, since these datasets come with test cases, we use the Pass@1 metric as the evaluation criterion along with the ASR metric on the poisoned dataset.

The results in Table~\ref{tab:larger-dataset} demonstrate the effectiveness of DeCE in mitigating backdoor attacks on large datasets.
We present the results of three different models on three different datasets, including the Clean model trained on the clean training set, the Poisoned model trained on the 10\% poisoned training set and the Defense model trained with DeCE on the 10\% poisoned training set. 
All models are trained in 2 epochs with a learning rate of 4e-5.
We observe that, when we poison the test set, the ASR of the model trained with DeCE is only 0\%-2\%, which indicates that DeCE is effective in defending against backdoor attacks on all three datasets. 
Notably, DeCE maintains a high Pass@1 score on the clean test set, indicating that it does not compromise the model's performance on clean data.

\subsection{Adaptive Attack}
In Section~\ref{sec:2.1}, we have introduced 
	data-poisoning backdoor attacks where attackers are assumed to be agnostic to the potential defence.  
For an adaptive attack where an attacker is aware 
	of the implementation of DeCE, they may design strategies to augment the concentration of poisoned samples within the dataset. 
	This presents a delicate balance for attackers. 
On the one hand, the increased percentage of poisoned samples may break the early learning phase of the model, thus increasing the likelihood of successful backdoor trigger insertion. 
On the other hand, 
a high percentage of poisoned instances may lead to noticeable irregularities in the dataset, which increases the likelihood of being detected by a vigilant user or a strong data integrity check.

\subsection{Threats to Validity}

In this section, we analyze potential threats to the validity of our empirical study.

\noindent\textbf{Threats to Internal Validity.}
The first internal threat is the possibility of implementation faults in DeCE. To mitigate this threat, we conduct a careful code inspection of the implementation and utilize well-established third-party libraries (such as PyTorch and Transformers).
The second internal threat is the implementation correctness of the considered baselines. To alleviate this threat, we implemented all baselines based on their shared models and scripts on platforms such as Hugging Face\footnote{\url{https://huggingface.co/models}} and Github.\footnote{\url{https://github.com}}

\noindent\textbf{Threats to External Validity.}
The main external threat lies in the datasets used in our study. To mitigate this threat, we carefully selected three high-quality datasets. 
For the code generation dataset, we select Lyra and Pisces, two high-quality
Turducken-style code datasets. Both datasets are collected through crowd-sourcing,
and each sample undergoes manual quality check  to ensure their
reliability and accuracy.
For the code repair dataset, we employ the Bugs2Fix dataset
from CodeXGLUE, which is a widely-adopted dataset within the
research community. 

\noindent\textbf{Threats to Construct Validity.}
The main construct threat is related to the metrics used in our automated evaluation. 
We first utilize the BLEU and CodeBLEU metric, where BLEU quantifies the token overlap between the synthesized code and reference implementations, and 
CodeBLEU is a variant of the BLEU metric accounting for the syntactic and semantic nuances of code.
To evaluate the effectiveness of backdoor attacks on poisoned
data, we introduce the ASR to measure the proportion of instances where the victim model, when presented with poisoned data containing specific triggers, produces the desired malicious output.
\section{Related Work}
\label{sec:related}

\subsection{Code Synthesis} 

In recent years, there have been significant advancements in the field of code synthesis~\cite{zan2023large}. Early approaches relied on expert systems  and domain-specific languages~\cite{liguori2021evil}, but they lacked flexibility and scalability. However, a recent surge in pre-trained language models (PLMs) based on the Transformer architecture~\cite{vaswani2017attention} has revolutionized code synthesis~\cite{ahmad2020transformer}.
These PLMs, trained on large-scale unlabeled code corpora, have performed remarkably in code synthesis tasks. They can be categorized into three groups: encoder-only (e.g., CodeBERT~\cite{feng2020codebert} and GraphCodeBERT~\cite{guo2020graphcodebert}), decoder-only (e.g., CodeGPT and CodeGPT-adapter~\cite{lu2021codexglue}), and enc-dec models (e.g., PLBART~\cite{ahmad2021unified}, CodeT5~\cite{wang2021codet5}, and NatGen~\cite{chakraborty2022natgen}).
In our task, we mainly focus on the enc-dec models which can combine the advantages of both encoder-only and decoder-only models, making them more suitable for generation tasks.

Furthermore, the development of large-scale pre-trained models with over 1 billion parameters (such as AlphaCode~\cite{li2022competition}, CodeGen~\cite{nijkamp2022codegen}, StarCoder~\cite{li2023starcoder}, CodeLlama~\cite{roziere2023code}, and CodeGeeX~\cite{zheng2023codegeex}) has further enhanced the performance of code synthesis. 

Different from the common focus on enhancing CLMs' performance on downstream tasks, our study emphasizes the security of these models, specifically tackling the threats of backdoor attacks.

\subsection{Backdoor Attack} 
Backdoor attacks pose a significant threat to neural network models, targeting the training phase rather than the inference phase, which can be classified into token-based, syntax-based, and semantic-based attacks in NLP.
Token-based attacks utilize trigger keywords to generate logical trigger sentences, while syntax-based attacks leverage syntactic triggers. For example, \citet{chen2021badnl} enhanced the effectiveness of token-based attacks by introducing semantic preservation trigger generation methods with multiple perturbation levels.
\citet{qi2021hidden} proposed a method that utilizes these triggers, and they also explored the use of text-style transfer techniques to generate more dynamic backdoor samples.
Semantic-based attacks focus on creating backdoor training samples that appear more natural to humans. \citet{chan2020poison} utilized an autoencoder to generate these samples, enhancing their authenticity.
Among these, token-based attacks demonstrate high attack efficiency but are more susceptible to detection. 
To overcome this limitation, \citet{chen2021badpre} proposed BadPre, a method that bypasses detection by randomly inserting triggers multiple times into the input sequence during deployment.
In the realm of programming languages, backdoor implantation has gained attention. Researchers have proposed various strategies, including fixed triggers~\cite{wan2022you}, rule-based poisoning~\cite{li2022poison}, and language model-guided poisoning~\cite{li2023multi}.
For instance, \citet{cotroneo2024vulnerabilities} proposed a data poisoning attack to assess the security of code generators by injecting software vulnerabilities to the training data.
\citet{sun2023backdooring} proposed a stealthy backdoor attack BADCODE against neural code search models by modifying variable/function names.

\subsection{Backdoor Defense}
Most studies defending against backdoor attacks have focused on models used in NLP.

\noindent\textbf{Active Defense.} Active defense methods aim to detect and remove backdoor samples defore or during the training phase. 
For example, \citet{chen2021mitigating} proposed Backdoor Keyword Identification (BKI), which identifies and removes potential poisoned samples during the training process. 
\citet{zhu2022moderate} proposed Moderate-fitting, which defends against backdoor attacks by reducing the model capacity, training epochs and learning rate.

\noindent\textbf{Passive Defense.} Passive defense methods aim to reduce the impact of backdoor attacks during the inference process. 
For example, \citet{qi2021onion} proposed ONION, a method that detects and removes discrete words in sentences using perplexity and output probability outputted by the language model. 
\citet{gao2021design} proposed STRIP, which detects and removes backdoor samples by analyzing the model's output. 
\citet{jain2023baseline} proposed a method that uses a generative model to interpret an adversarial instruction. Ideally, the generative model will accurately retain the natural instruction and may remove the malicious trigger in the instruction. Although the interpretation instruction works well in most cases, it may also cause model degradation.



In our study, we focus on developing active defense methods against backdoor attacks. 
Our defense method leverages the "early learning" phenomenon observed during the training of CLMs. 
Our proposed method not only showcases enhanced effectiveness but also exhibits a wider applicability scope when compared with previous defense methodologies.
\section{Conclusion}
\label{sec:conclusion}
In this study, we reproduce the "early learning" phenomenon in CLMs and propose DeCE that mitigates the impact of backdoor triggers on model behavior. Through extensive experiments on multiple code synthesis datasets, models, and poisoning ratios, we demonstrate the effectiveness of DeCE in defending against backdoor attacks.

While DeCE has shown promising results in defending against backdoor attacks, we 
want to optimize its hyper-parameters in the future, which can improve the defense quality against various attack strategies.
Futhermore, we will explore more covert and complex poisoning attack methods to thoroughly evaluate the proposed defense mechanism's performance in the real world.
On the other hand, we would like to study its applicability in more areas of code intelligence and NLP, such as text classification, code summarization and other tasks.

\begin{acks}
This research/project is supported by the National Natural Science Foundation of China (No. 62372232), the Short-term Visiting Program of Nanjing University of Aeronautics and Astronautics for Ph.D. Students Abroad (No. 240602DF16), High Performance Computing Platform of Nanjing University of Aeronautics and Astronautics, and the Collaborative Innovation Center of Novel Software Technology and Industrialization. 
T. Chen is partially supported by an oversea grant from the State Key Laboratory of Novel Software Technology, Nanjing University (KFKT2022A03, KFKT2023A04). 
This research/project is supported by the National Research Foundation, under its Investigatorship Grant (NRF-NRFI08-2022-0002). Any opinions, findings and conclusions or recommendations expressed in this material are those of the author(s) and do not reflect the views of National Research Foundation, Singapore.
\end{acks}
 
\normalem
\bibliographystyle{Reference}
\bibliography{acmart}
 
\end{document}